\RequirePackage{fix-cm}

\newcommand \be{\begin{equation}}
\newcommand \en{\end{equation}}
\newcommand \bea{\begin{eqnarray}}
\newcommand \ena{\end{eqnarray}}

\documentclass[smallextended]{svjour3}       
\smartqed  
\usepackage{graphicx}
\usepackage{dcolumn}
\usepackage{bm}
\usepackage{ifpdf}
\usepackage{hyperref}
\usepackage{dcolumn}
\usepackage{bm}
\usepackage[spanish,english]{babel}
\usepackage{amsfonts}
\usepackage{amssymb}
\usepackage{graphicx}
\usepackage[latin1]{inputenc}

 \journalname{GRG}
\begin{document}

\title{Thermodynamic analysis of black hole solutions in gravitating nonlinear electrodynamics
}

\titlerunning{Thermodynamic analysis of black hole solutions in gravitating nonlinear...}        

\author{J. Diaz-Alonso         \and
        D. Rubiera-Garcia
}

\institute{J. Diaz-Alonso \and D. Rubiera-Garcia \at
              LUTH, Observatoire de Paris, CNRS, Universit\'e Paris
Diderot. 5 Place Jules Janssen, 92190 Meudon, France.
\\ Departamento de F\'isica, Universidad de Oviedo. Avenida
Calvo Sotelo 18, 33007 Oviedo, Asturias, Spain.\\
              \email{joaquin.diaz@obspm.fr}           
           \\
            \email{drubiera@fisica.ufpb.br} \and
            \emph{present address} \\
            D. Rubiera-Garcia \\ 
Departamento de F\'isica, Universidade Federal da Para\'iba, Jo\~ao Pessoa, Para\'ba 58051-900, Brazil.
}

\date{Received: date / Accepted: date}

\maketitle

\begin{abstract}
We perform a general study of the thermodynamic properties of static electrically charged black hole solutions of nonlinear electrodynamics minimally coupled to gravitation in three space dimensions. The Lagrangian densities governing the dynamics of these models in flat space are defined as arbitrary functions of the gauge field invariants, constrained by some requirements for physical admissibility. The exhaustive classification of these theories in flat space, in terms of the behaviour of the Lagrangian densities in vacuum and on the boundary of their domain of definition, defines twelve families of admissible models. When these models are coupled to gravity, the flat space classification leads to a complete characterization of the associated sets of gravitating electrostatic spherically symmetric solutions by their central and asymptotic behaviours. We focus on nine of these families, which support asymptotically Schwarzschild-like black hole configurations, for which the thermodynamic analysis is possible and pertinent. In this way, the thermodynamic laws are extended to the sets of black hole solutions of these families, for which the generic behaviours of the relevant state variables are classified and thoroughly analyzed in terms of the aforementioned boundary properties of the Lagrangians. Moreover, we find universal scaling laws (which hold and are the same for all the black hole solutions of models belonging to any of the nine families) running the thermodynamic variables with the electric charge and the horizon radius. These scale transformations form a one-parameter multiplicative group, leading to universal ``renormalization group"-like first-order differential equations. The beams of characteristics of these equations generate the full set of black hole states associated to any of these gravitating nonlinear electrodynamics. Moreover the application of the scaling laws allows to find a universal finite relation between the thermodynamic variables, which is seen as a generalized Smarr law. Some particular well known (and also other new) models are analyzed as illustrative examples of these procedures.

\end{abstract}

\section{Introduction} \label{sectionI}

In General Relativity the theorems on singularities, on the uniqueness of the black hole solutions of the Einstein vacuum field equations, and the Cosmic Censorship conjecture \cite{carter71} lead to the strong belief that the Kerr spacetime describes accurately the outcome of the gravitational collapse of rotating electrically neutral matter distributions. When such distributions support a non-vanishing total electric charge, similar arguments, now involving the minimally coupled Einstein-Maxwell theory, lead to Kerr-Newman black hole configurations as the final state of the collapse \cite{MTW}.

However, there are several arguments justifying the study of other BH configurations \footnote{In order to facilitate the reading, let us give a list of the acronyms used throughout this paper: \textbf{NED} ($\equiv$ nonlinear electrodynamics), \textbf{G-NED} ($\equiv$ gravitating-nonlinear electrodynamics), \textbf{ADM} ($\equiv$ Arnowitt-Deser-Misner), \textbf{BH} ($\equiv$ black holes), \textbf{BP} ($\equiv$ black points), \textbf{RN} ($\equiv$ Reissner-Nordstr\"om), \textbf{BI} ($\equiv$ Born-Infeld), \textbf{EH} ($\equiv$ Euler-Heisenberg), \textbf{ESS} ($\equiv$ electrostatic spherically symmetric), \textbf{G-ESS} ($\equiv$ gravitating electrostatic spherically symmetric), \textbf{UVD} ($\equiv$ ultraviolet divergent), \textbf{IRD} ($\equiv$ infrared divergent).}, which are solutions of more complex gravitating NEDs, as possible final states of the gravitational collapse of electrically charged matter. A first argument comes from the low energy regime of string theory, where fundamental excitation is gravity. If one accepts the fundamental character of this theory, then the dynamics of the emerging gauge fields in this regime is governed by non-minimal effective Lagrangians \cite{ST}. In this context some nonlinear electrodynamics, as the Born-Infeld model \cite{BI34}, are to be seen as more ``fundamental", in describing the dynamics of the electromagnetic fields, than the usual Maxwell theory, which is regarded, within this picture, as a weak field approximation. Besides that, Born-Infeld generalizations of SU(2) non-Abelian gauge field theory, coupled to tensor-scalar gravity, have been also used for the description of dark energy in a cosmological context \cite{darkenerg}.

A second argument concerns the quantum vacuum effects on the electromagnetic field, which are neglected in the classical Einstein-Maxwell theory. Now accepting the fundamental character of the Maxwell theory and its quantum formulation, the corrections of the Dirac vacuum on the dynamics of electromagnetic fields in flat space can be described, at a phenomenological classical level, through nonlinear effective Lagrangians that define NEDs, the first historical example being the EH Lagrangian \cite{EH36}. It accounts for the nonlinear effects of the Dirac vacuum on the low energy electromagnetic wave propagation, calculated to lowest order in the perturbative expansion. When higher order operators are included in the perturbative calculation we are led to a sequence of effective Lagrangians which take the form of polynomials in the field invariants, arranged as an expansion in operators of increasing dimensions \cite{birula70}. These effective Lagrangians exhibit finite-energy elementary solutions in flat space, even though the bare (Maxwell) Lagrangian does not \cite{dr09}. When minimally coupled to gravity such Lagrangians describe satisfactorily the vacuum effects in the slowly varying curvature regions \cite{dobado97}. Disregarding the rotational degrees of freedom, which are not essential for the present considerations, the elementary solutions of any physically \textit{admissible} NED (a concept to be defined below) minimally coupled to gravity, are asymptotically flat, exhibit a curvature singularity at the center and their electrostatic field strengths are monotonically decreasing functions of the radius \cite{dr10a,dr10b}. Near the singularity this effective approach fails, and the description of the vacuum effects would require a consistent quantum treatment of both the electromagnetic and gravitational fields, which is not available at present. But in moving away from the singularity the curvature stabilizes very quickly and becomes slowly varying in the characteristic distances of the vacuum screening. Consequently, the effective approach remains accurate everywhere in the manifold, excepted in a small region close to the singularity. Therefore, in describing the collapsed state of non-rotating electrically charged matter, the BH configurations associated to some NEDs (as EH) could be more ``realistic" than the Reissner-Nordstr\"om one.

A general analysis of the gravitational configurations and geometric structures of the elementary solutions associated to the whole class of admissible NEDs, minimally coupled to gravity (including some well known models such as Maxwell, Born-Infeld or Euler-Heisenberg, as particular representative examples) has been performed in Refs.\cite{dr10a,dr10b}. Nevertheless, the issue of the thermodynamic properties and laws of the corresponding BH solutions was not considered in those works. Such a study is the main object of this paper.

The thermodynamic analogies for asymptotically flat BH solutions of the Einstein equations were established in the early seventies for the cases of the Kerr-Newman BH and its non-rotating and uncharged limits \cite{varios}. The subsequent discovery of Hawking's radiation \cite{Hawking} turned these analogies into actual laws of the thermodynamics of these configurations. Since then, an important amount of literature has dealt with this issue in depth. In this way many papers have analyzed the thermodynamics of the Reissner-Nordstr\"om black hole (see e.g. \cite{RN-thermo} and references therein, also \cite{chamblin99}). Let us mention some extensions and developments going beyond the analysis of this solution: i) the validity of the thermodynamic laws has been established for asymptotically flat BH solutions of several classes of field theories minimally coupled to gravity, including NEDs supporting asymptotically coulombian elementary solutions \cite{rasheed97}. In this way the thermodynamic analysis has been performed for asymptotically flat BH solutions associated to some particular NEDs, as the gravitating BI model \cite{BI-thermo} or, more recently, the family of NED models defined through Lagrangian densities which are constructed as powers of the Maxwell one \cite{hassaine09-NED}; ii) this analysis has been also performed for BH solutions of the gravitating Born-Infeld NED in asymptotically cosmological backgrounds, as (anti-)de Sitter spaces in ($3+1$) dimensions \cite{bhads,BI-AdS}, in higher dimensions \cite{BI-AdS-higher,bhads2} and in ($2+1$) dimensions \cite{BI2}, extending the Hawking-Page results of the uncharged case \cite{Hawking83} and motivated by the AdS/CFT correspondence \cite{adscft}; iii) several extensions of these solutions, resulting from the minimal coupling of NEDs to gravitational dynamics defined beyond the Einstein-Hilbert action, have been considered, as in the cases of Gauss-Bonnet and Lovelock \cite{aiello}, $f(R)$ \cite{fR} or scalar-tensor theories \cite{doneva10}; iv) statistical computations of the entropy for the extreme and near extreme five dimensional RN black holes have been performed in the framework of string theory \cite{strominger96}.

However, most of the aforementioned works deal with some particular models, while a comprehensive study of the thermodynamic properties of BH configurations associated to general gravitating NEDs is possible, but still lacking. The aim of this paper is to provide such a study, by analyzing the full set of NEDs minimally coupled to gravitation in ($3+1$) dimensions and defined by Lagrangian densities which are arbitrary functions of the two field invariants, but restricted by some consistency requirements (such as the positivity of the energy, the regularity of the Lagrangian and the parity invariance) leading to what we shall call \emph{admissible} theories. In Ref.\cite{dr10b} the gravitating admissible NEDs were classified into two sets, according to the form of the asymptotic behaviours of their elementary solutions. These solutions, which \underline{are always asymptotically flat}, can approach flatness as the Schwarzschild field (\emph{normal} case) or slower than it (\emph{anomalous} case). In the anomalous case some thermodynamic variables (such as the ADM mass) cannot be defined and the thermodynamic analysis is not possible, at least in the usual way. Consequently, in the present work we shall deal only with the analysis of the thermodynamics of asymptotically normal BH solutions supported by the corresponding families of G-NEDs.

In section \ref{sectionII} we summarize some results of Ref.\cite{dr09} regarding the elementary solutions of NEDs in flat space, which are necessary for the present study. In particular, we introduce the classification of these theories that, when generalized to the gravitating versions, defines the different families of possible BH solutions.

In section \ref{sectionIII} we briefly describe the geometrical structures characterizing the BHs associated to each family. This will be necessary in the next section for a clear understanding of the thermodynamical behaviours. A detailed report of the contents of this section has been published in Ref.\cite{dr10a}.

Section \ref{sectionIV} is devoted to the thermodynamical analysis itself. First, we prove the validity of the expression of the first law for any admissible gravitating NED supporting \underline{asymptotically normal} BH solutions. In particular this validity is established for those models with non Maxwellian weak field limit, for which the theorems of Ref.\cite{rasheed97} do not immediately apply. Next we analyze the behaviour of the main thermodynamical variables and specific heats for the BH configurations associated to the different families. In particular the study of the thermal properties of the extreme BH configurations suggests the possible presence of van der Waals-like first-order phase transitions between these states for models belonging to several families. The results of this analysis are illustrated with some well known examples (as the RN, EH and BI) representative of different families, while other examples are specifically built to exhibit some particular behaviours.

In section \ref{sectionV} we find some \textbf{universal} relations between the thermodynamic variables of the asymptotically-normal BH states associated to admissible G-NEDs. First, we obtain a universal expression for the \textit{generalized Smarr law}, valid for all such BH solutions. This law reduces to the usual Smarr formula in the RN case and to the corresponding expressions obtained for the few particular examples of G-NEDs studied in the literature. Next, we obtain scaling laws satisfied by the functions relating the thermodynamic variables, which are \textit{universal} (i.e. valid for any G-NED). We make a preliminary discussion on some implications of these laws. In particular, we show the existence of a one-parameter multiplicative group structure underlying the scaling transformations and obtain the associated first-order differential equations (``renormalization group"-like equations) relating the derivatives of several state variables with respect to the charge ($Q$) and the horizon radius ($r_{h}$) of the BH solutions. These equations are the same for any G-NED and the associated \textit{characteristic} curves in the corresponding three-dimensional space define the running of the thermodynamic variables and generate the ``equations of state" for the sets of BH solutions associated to any gravitating NED (defined as surfaces supported by the bean of characteristics).

Finally, we conclude in section \ref{sectionVI} with some comments and perspectives.

\section{General results on NEDs in flat space-time} \label{sectionII}

Let us summarize the results concerning the complete characterization and classification of the whole set of admissible NEDs through the properties of their ESS solutions \cite{dr09}, which will be necessary for the analysis of the gravitating problem. The dynamics of these fields in flat four-dimensional space-time are governed by Lagrangian densities which are single-branched functions

\be
\varphi(X,Y),
\label{eq:(2-1)}
\en
of the two gauge field invariants $X = -\frac{1}{2} F_{\mu\nu}F^{\mu\nu} = \vec{E}^{2} - \vec{H}^{2}$ and $Y = -\frac{1}{2} F_{\mu\nu}F^{*\mu\nu} = 2\vec{E}\cdot\vec{H}$, where $F_{\mu\nu}=\partial_{\mu}A_{\nu}-\partial_{\nu}A_{\mu}$ is the field strength tensor of the vector potential $A_{\mu}$, $F^{*\mu\nu}=\frac{1}{2}\epsilon^{\mu\nu\alpha\beta}F_{\alpha\beta}$ its dual and $\vec{E}$ and $\vec{H}$ the electric and magnetic fields, respectively. For physical consistence of the corresponding theories these functions must be defined in a open and connected domain of the $X-Y$ plane, including the vacuum, and be restricted by the conditions of parity invariance ($\varphi(X,Y)=\varphi(X,-Y)$) and positivity of the energy. The latter condition reads \cite{dr09}

\be
\rho \geq \left(\sqrt{X^{2}+Y^{2}} + X\right) \frac{\partial \varphi}{\partial X}+ Y\frac{\partial \varphi}{\partial Y} - \varphi(X,Y) \geq 0,
\label{eq:(2-2)}
\en
with vanishing vacuum energy ($\varphi(0,0) = 0$). Moreover, the ESS solutions $E(r)$ must be single-branched, defined and regular for $r>0$. This requires $\varphi(X,0)$ to be a strictly monotonically increasing function of $X$ in the range of values $X=E^{2}(r>0)$ ($\partial \varphi/\partial X\mid_{Y=0}$ is required to be continuous and strictly positive there). The above set of requirements define what we have called admissible NEDs, the only ones that we shall consider here. In addition one can assume $\varphi(X,Y)$ to satisfy the condition

\be
\frac{\partial \varphi}{\partial X} \geq 2X\frac{\partial^{2} \varphi}{\partial Y^{2}},
\label{eq:(2-3)}
\en
(for $X=E^{2}(r>0), Y=0$) which is \textbf{necessary and sufficient} for the linear stability of the ESS solutions, as shown in Ref.\cite{dr09}.

The Euler field equations associated to the Lagrangians (\ref{eq:(2-1)}) read

\be
\partial_{\mu} \left[\frac{\partial \varphi}{\partial X} F^{\mu\nu} + \frac{\partial \varphi}{\partial Y} F^{*\mu\nu}\right] = 0,
\label{eq:(2-3)bis}
\en
and the corresponding energy-momentum tensor takes the form

\be
T_{\mu\nu} = 2F_{\mu\alpha}\left(\frac{\partial \varphi}{\partial X}F_{\nu}^{\alpha} + \frac{\partial \varphi}{\partial Y}F_{\nu}^{*\alpha}\right) - \varphi \eta_{\mu\nu}.
\label{eq:(2-3)ter}
\en
If we look for ESS solutions of the field equations ($\vec{E}(\vec{r}) = E(r)\vec{r}/r$, $\vec{H}=0$) we find the first integral

\be
r^{2}\frac{\partial \varphi}{\partial X}\Big \vert_{Y=0} E(r) = Q,
\label{eq:(2-4)}
\en
where the integration constant $Q$ is identified with the point-like charge source of the solution. Once the form of $\varphi(X,Y)$ is specified Eq.(\ref{eq:(2-4)}) gives in implicit form the expression of the ESS field $E(r,Q)$ as a function of $R = r/\sqrt{ \vert Q\vert }$, in such a way that it scales as

\be
\vert E(r,Q) \vert = \vert E(R,Q = 1) \vert.
\label{eq:(2-4)bis}
\en
As we shall see below, this scaling law induces similar laws in the different thermodynamic functions describing the systems of BH solutions of any gravitating NED. We shall call ``characteristic" the $Q=1$ configurations, from which the whole thermodynamic structure of the set of BH solutions associated to a given G-NED is determined through these scaling transformations (see sections \ref{sectionIV} and \ref{sectionV} below).

The positivity of the derivative of the Lagrangian function in Eq.(\ref{eq:(2-4)}) allows to restrict the analysis to the case ($E>0, Q>0$) without loss of generality. Assuming behaviours for the ESS solutions around the center and asymptotically of the form

\be
E(r \rightarrow 0,Q) \sim \nu_{1}(Q)r^{p} \hspace{0.3cm};\hspace{0.3cm} E(r \rightarrow \infty,Q) \sim \nu_{2}(Q)r^{q},
\label{eq:(2-5)}
\en
the admissibility conditions restrict the values of the exponents to $p \leq 0$ and $q < 0$. At the center the fields diverge for $p<0$ and behave as

\be
E(r \rightarrow 0) \sim a - b(Q) r^{\sigma},
\label{eq:(2-6)}
\en
for $p=0$. In the latter case the parameter $a$ (the maximum field strength) and the exponent $\sigma > 0$ are universal constants for a given model, whereas the coefficient $b(Q)$ is related to the charge of each particular solution as

\be
b(Q) Q^{\sigma/2} = \lim_{X \rightarrow a^{2}} (a - \sqrt{X}) \left[a \frac{\partial \varphi}{\partial X}\right]^{\sigma/2} = b_{0},
\label{eq:(2-7)}
\en
$b_{0} = b(Q = 1)$ being also a universal constant of the model. Asymptotically $E(r \rightarrow \infty)$ vanishes for the ESS solutions of any admissible model.

The corresponding behaviours of the Lagrangian densities around these values of the fields ($X = E^{2}(r)$) are given by the expression

\be
\varphi(X,Y=0) \sim \alpha_{i} X^{\gamma_{i}},
\label{eq:(2-8)}
\en
where $\alpha_{i}, (i=1,2)$ are positive constants which are related with the coefficients in Eq.(\ref{eq:(2-5)}). For $X \rightarrow \infty$ (and $p < 0$) we have

\be
\nu_{1}(Q) = \left(\frac{\gamma_{1}\alpha_{1}}{Q}\right)^{p/2}; \hspace{.2cm} \gamma_{1} = \frac{p-2}{2p},
\label{eq:(2-9)}
\en
whereas for $X \rightarrow 0$ we have

\be
\nu_{2}(Q) = \left(\frac{\gamma_{2}\alpha_{2}}{Q}\right)^{q/2}; \hspace{.2cm} \gamma_{2} = \frac{q-2}{2q}.
\label{eq:(2-10)}
\en
In both cases $\gamma_{i}>1/2$. If $p = 0$ the Lagrangian densities behave around the center ($X = E^{2}(r=0) = a^{2}, Y=0$) as

\be
\varphi(X,Y=0) \sim \frac{2 \sigma b_{0}^{2/\sigma}}{2-\sigma}(a - \sqrt{X})^{\frac{\sigma-2}{\sigma}} + \Delta,
\label{eq:(2-11)}
\en
if $\sigma \neq 2$. If $\sigma = 2$ this behaviour becomes

\be
\varphi(X,Y=0) \sim - b_{0} \ln(a - \sqrt{X}) + \Delta.
\label{eq:(2-12)}
\en
In these equations $\Delta = \varphi(a^{2},Y=0) < \infty$ if $\sigma > 2,$ and it is a universal constant of the model. If $\sigma \leq 2$ the value of $\Delta$ can be calculated from the explicit expression of the Lagrangian density (see Eqs.(\ref{eq:(4-18)bis}) and (\ref{eq:(4-18)ter}) below). Consequently, for $\sigma \leq 2$ the Lagrangians exhibit a vertical asymptote at $X=a^{2}$. For $\sigma > 2$ they take finite values there, with divergent slope (the Born-Infeld model \cite{BI34} belongs to this case with $\sigma = 4$).

Equation (\ref{eq:(2-3)ter}) leads to the expression

\be
\rho = T_{0}^{0} = 2\frac{\partial \varphi}{\partial X}\Big \vert_{Y=0} \vec{E}^{2}(r,Q) - \varphi(X,Y) = \frac{2QE - r^{2}\varphi}{r^{2}}
\label{eq:(2-13)}
\en
for the energy density distribution of the ESS solutions. As easily seen, the spatial integrals of these functions around the center of the ESS solutions converge if $-1 < p \leq 0$ and diverge if $p \leq -1$. This allows a classification of the admissible NEDs in terms of the central behaviour of their ESS solutions, leading to three families. We have called \textbf{A1} the first family, corresponding to the range of values $-1 < p < 0$, for which the ESS solutions diverge as $r \rightarrow 0$, but the integrals of energy converge there. For these models the Lagrangian functions behave, when $X \rightarrow \infty$, as in Eq.(\ref{eq:(2-8)}) with $\gamma > 3/2$. We have called \textbf{A2} the second family, for which $p = 0$, corresponding to the NEDs supporting ESS solutions which are finite at the center. Obviously, the integrals of energy converge there for these solutions (see Eqs.(\ref{eq:(2-11)})-(\ref{eq:(2-13)})). The third family corresponds to $p \leq -1$ and supports ESS solutions diverging at the center and whose integrals of energy diverge also there (we have called these models \textbf{UVD}, acronym for ultraviolet divergent). When the integral of energy is convergent at the center (cases \textbf{A1} and \textbf{A2}) the ``internal energy function"

\be
\varepsilon_{in}(r,Q) = 4\pi \int_{0}^{r} R^{2} \rho(R,Q) dR,
\label{eq:(2-14)}
\en
giving the energy of the ESS field of a point-like charge $Q$ inside a sphere of radius $r$, is well defined.

The asymptotic behaviour of the ESS solutions allows for a second classification of the admissible NEDs. For large $r$ the integrals of energy converge asymptotically if $q < -1$ and diverge if $-1 \leq q < 0$. Thus we have defined four classes of models differing by this behaviour: \textbf{B1} models (when $-2 < q < -1$) whose ESS solutions are asymptotically damped slower than the Coulomb field and the integrals of energy converge asymptotically; \textbf{B2} models (when $q = -2$) whose ESS solutions behave asymptotically as the Coulomb field; \textbf{B3} models (when $q < -2$) whose ESS solutions vanish asymptotically faster than the Coulomb field; \textbf{IRD} models (acronym for infrared divergent) when $-1 \leq q < 0$, whose ESS solutions are asymptotically damped slower than the Coulomb field and their integrals of energy diverge asymptotically. When the integral of energy converges asymptotically (cases \textbf{B1}, \textbf{B2} and \textbf{B3}) the ``external energy function"

\be
\varepsilon_{ex}(r,Q) = 4\pi \int_{r}^{\infty} R^{2} \rho(R,Q) dR,
\label{eq:(2-15)}
\en
giving the energy of the field outside the sphere of radius $r$, is well defined.

The combination of these central and asymptotic behaviours allows for the exhaustive classification of the admissible NEDs in twelve classes. Six of them (combinations of \textbf{A1} and \textbf{A2} with \textbf{B1}, \textbf{B2} and \textbf{B3} behaviours) support finite-energy ESS solutions, which are to be interpreted as non-topological solitons if the necessary and sufficient stability condition (\ref{eq:(2-3)}) is fulfilled by the corresponding Lagrangian densities in the domain of definition of these solutions \cite{dr09}. For these six families the convergence of the integral of energy in the whole space leads to the relations

\be
\varepsilon(Q) = \varepsilon_{in}(\infty,Q) = \varepsilon_{ex}(0,Q) = \varepsilon_{in}(r,Q) + \varepsilon_{ex}(r,Q)
\label{eq:(2-16)}
\en
for the energy functions associated to the ESS solutions of charge $Q$. Using Eqs.(\ref{eq:(2-4)}) and (\ref{eq:(2-13)}) it can be easily seen that these energies, when they are well defined, scale as

\bea \label{eq:(2-16)bis}
\varepsilon(Q) &=& Q^{3/2}\varepsilon(Q = 1); \nonumber \\ \varepsilon_{in}(r,Q) &=& Q^{3/2}\varepsilon_{in}\left(\frac{r}{\sqrt{Q}},Q = 1 \right); \\ \varepsilon_{ex}(r,Q) &=& Q^{3/2}\varepsilon_{ex}\left(\frac{r}{\sqrt{Q}},Q = 1 \right), \nonumber
\ena
where the total energy of the characteristic configuration $\varepsilon(Q = 1)$ is a universal constant for a given model. By using the last expression  of $\varrho$ in Eq.(\ref{eq:(2-13)}) for the calculation of (\ref{eq:(2-15)}) as $r \rightarrow 0$, after some partial integrations and taking account the first integral (\ref{eq:(2-4)}), as well as the central and asymptotic behaviours of the fields in these finite-energy cases, we are led to the useful expression

\be
\varepsilon(Q) = \frac{16\pi Q}{3} \int_{0}^{\infty} E(R,Q) dR,
\label{eq:(2-16)ter}
\en
relating the total energy of the soliton with the integral of the field.

The ESS solutions of the remaining six cases (combinations including \textbf{UVD, IRD} or both behaviours) are energy divergent. When coupled to gravity all the models of the twelve families lead to consistent gravitating field theories. However, we shall see that the families involving the \textbf{IRD} behaviour fail to allow for a thermodynamic analysis in the usual way.

\begin{figure}[h]
 \begin{center}
\includegraphics[width=0.75\textwidth]{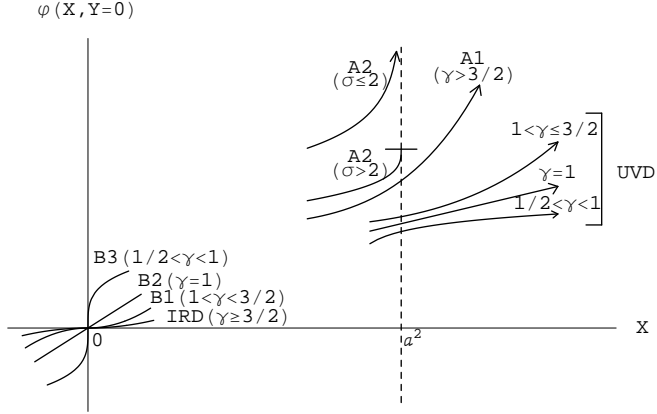}
\caption{Qualitative behaviour of the admissible Lagrangian densities $\varphi(X,Y=0)$ i) for small ESS fields ($E^{2} = X \sim 0,$ corresponding to the \textbf{B} and \textbf{IRD} asymptotic behaviours of the ESS solutions), ii) for large ESS fields ($E^{2} = X \rightarrow \infty,$ corresponding to the \textbf{A1} and \textbf{UVD} central-field behaviours) and iii) for finite maximum field-strength models ($X = E^{2}_{max} = a^{2}$), corresponding to the \textbf{A2} central-field behaviour. The Lagrangian densities behave as $X^{\gamma_{2}}$ around $X=0$ (in the \textbf{B} and \textbf{IRD} cases) and as $X^{\gamma_{1}}$ around $X \rightarrow \infty$ (in the \textbf{A1} and \textbf{UVD} cases) where the $\gamma_{i}$ constants are related to the central and asymptotic behaviours of the ESS solutions through Eqs.(\ref{eq:(2-9)}) and (\ref{eq:(2-10)}) respectively. In the \textbf{A2} cases the Lagrangian density exhibits a vertical asymptote at $X = a^{2}$ (if $\sigma \leq 2$) or takes a finite value with divergent slope there (if $\sigma > 2$). In the intermediate range of $X$ values, matching the central and asymptotic regions, $\varphi(X,Y=0)$ must be strictly monotonically increasing, for admissibility.}
\label{figure1}
 \end{center}
\end{figure}

In figure \ref{figure1} we have plotted the profiles of the Lagrangian density functions of the admissible NEDs (for $Y=0$) around the vacuum and on the boundary of their domain of definition.

\section{General results on gravitating NEDs} \label{sectionIII}

The full set of admissible NEDs minimally coupled to gravity, and the structure of their G-ESS solutions have been classified and extensively analyzed in Refs.\cite{dr10a,dr10b}. However, as mentioned in section \ref{sectionI}, this analysis concerned mainly the geometric structure of these solutions and excluded the study of their thermodynamic properties. Before tackling this thermodynamic problem let us summarize some pertinent results of these references.

As usual, the action corresponding to the minimal coupling of gravity and any NED reads

\be
\textit{S} = \textit{S}_{G} + \textit{S}_{NED} = \int d^4x \sqrt{-g}\left[\frac{R}{16\pi G} - \varphi(X,Y)\right],
\label{eq:(3-1)}
\en
$g$ and $R$ being the metric determinant and the curvature scalar, respectively. The associated Einstein equations for the G-ESS solutions can be written in an adapted Schwarzschild-like coordinate system, with an interval of the form

\be
ds^{2} = g(r) dt^{2} - \frac{dr^{2}}{g(r)} - r^{2} d\Omega^{2},
\label{eq:(3-2)}
\en
where $d\Omega^{2}=d\theta^2 +\sin^2 (\theta) d \psi^2$ is the angular metric and $g_{00} = g(r)$ is the only metric function to be determined. The simplicity of the form of this interval is a consequence of the Einstein equations together with the equality of the \textit{mixed} components of the energy-momentum tensor, $T_0^0=T_1^1$, which remains the same as in flat space in spherical coordinates. These equations reduce to

\bea \label{eq:(3-3)}
\frac{d}{dr}\left(r g(r) - r\right) &=& -8\pi r^{2} T^{0}_{0} = -8\pi r^{2}\left(2\frac{\partial \varphi}{\partial X} E^{2} - \varphi\right)\nonumber \\
\frac{d^{2}}{dr^{2}}\left(r g(r)\right) &=& -16\pi r T^{2}_{2} = 16\pi r \varphi,
\ena
whose compatibility can be easily established. The general solution of this system reads

\be
g(r,Q,C) = 1 + \frac{2C}{r} - \frac{8\pi}{r}\int r^{2} T^{0}_{0}(r,Q) dr,
\label{eq:(3-4)}
\en
where $C$ is an arbitrary integration constant. The characterization and classification of the admissible NEDs in flat space, summarized in section \ref{sectionII}, allows for the complete characterization and classification of the corresponding G-NEDs. The asymptotic behaviour of the metric function is dominated by the behaviour at large $r$ of the ESS solutions in flat space and is independent of their properties at finite $r$. Indeed, when the asymptotic behaviour of the ESS solutions in flat space belongs to the cases \textbf{B1}, \textbf{B2} or \textbf{B3} the metric function can be written as

\be
g(r,Q,M) = 1 - \frac{2M}{r} + \frac{2\varepsilon_{ex}(r,Q)}{r},
\label{eq:(3-5)}
\en
where $\varepsilon_{ex}(r,Q)$ is the external energy function, defined by (\ref{eq:(2-15)}), and the constant $M$ is identified as the ADM mass. Owing to the asymptotic behaviour of the external energy function in these cases ($\sim r^{q+1}, q < -1$), the last term in (\ref{eq:(3-5)}) can be neglected at large $r$ and the metric function behaves asymptotically as the Schwarzschild one \cite{dr10a}. The combinations of these asymptotic \textbf{B}-cases with the central-field \textbf{A1} and \textbf{A2} cases, which support finite-energy ESS solutions in flat space, admit an alternative expression for the metric function in terms of the internal energy function, which reads

\be
g(r,Q,M) = 1 - \frac{2U}{r} - \frac{2\varepsilon_{in}(r,Q)}{r},
\label{eq:(3-6)}
\en
where the constant $U$ is now related to the gravitational mass and the electrostatic soliton energy in flat space through

\be
U = M - \varepsilon(Q).
\label{eq:(3-7)}
\en
This constant may be roughly interpreted as the gravitational binding energy of the configuration plus the proper energy of the point-like source (see, however, the discussion about this kind of interpretations and the associated difficulties in Ref.\cite{ortin} and the pertinent references therein). In reference \cite{dr10a} we have called ``critical" the configurations with $U = 0$. As we shall see in the next section, in the finite field amplitude cases (\textbf{A2}) with the particular value of the charge

\be
Q_{c} = \frac{1}{16\pi a},
\label{eq:(3-7)bis}
\en
the critical configurations correspond to extreme black points \cite{Soleng95}, which exhibit a rich and peculiar thermodynamical behaviour. With $Q < Q_{c}$ they correspond to non-extreme BPs. In \textbf{A2} cases with $Q > Q_{c}$ (or for \textbf{A1} cases with any $Q$), the $U = 0$ configurations correspond to BH with an external event horizon and a vanishing-radius Cauchy inner horizon.

For models belonging asymptotically to \textbf{B}-cases and with \textbf{UVD} central-field behaviour (e.g. Maxwell Lagrangian) the electrostatic energy diverges and only the expression (\ref{eq:(3-5)}) for the metric function remains valid. For NEDs with \textbf{IRD} solutions in flat space the G-ESS solutions are asymptotically flat, but not Schwarzschild-like (asymptotically anomalous behaviour) and the ADM mass cannot be defined \cite{dr10b}. In these cases the metric function takes the form (\ref{eq:(3-6)}) if the central-field behaviour belongs to cases \textbf{A1} or \textbf{A2} (but now the interpretation (\ref{eq:(3-7)}) of the constant $U$ makes no sense). For the families with \textbf{UVD-IRD}  behaviour, only the expression (\ref{eq:(3-4)}) describes properly the form of the metrics, which are thermodynamically meaningless.

At finite $r$ the geometric structure of the G-ESS solutions is characterized by the central behaviour of the electric field in flat space. The event horizons (if any) are located on the largest solution of the equation $g(r) = 0$. For models with \textbf{UVD} central-behaviour solutions the metrics exhibit a structure similar to that of the Reissner-Nordstr\"om solution (naked singularities, extreme black holes or two-horizons black holes). If the central behaviour belongs to cases \textbf{A1} or \textbf{A2} we have, in addition, new kinds of solutions: non-extreme single-horizon black holes (for families \textbf{A1} and \textbf{A2}) and extreme and non-extreme black points (for families \textbf{A2}) \cite{dr10a}.

All the G-ESS solutions of the Einstein equations minimally coupled to admissible NEDs must belong to one of the twelve classes obtained as combinations of these asymptotic and central-field behaviours. Consequently, these results, which are summarized in table \ref{table:I}, characterize qualitatively the possible geometric structures of the G-ESS solutions of all admissible G-NEDs.

\begin{table}
 \begin{center}
   \begin{tabular}{| c | c | c | }
        \hline
      & B1, B2, B3  & IRD  \\ \hline
     & FEFS & DEFS \\
      A1& AS & AA \\
      & NS, EBH, 2hBH, 1hBH & NS, EBH, 2hBH, 1hBH \\ \hline
      & FEFS & DEFS \\
       A2 & AS & AA \\
      &  NS, EBH, 2hBH,1hBH, & NS, EBH, 2hBH, 1hBH, \\
      &  BP, EBP& BP, EBP \\  \hline
      & DEFS & DEFS \\
      UVD & AS & AA \\
      & NS, EBH, 2hBH & NS, EBH, 2hBH \\
     \hline
   \end{tabular}
 \caption{Possible geometric structures of the G-ESS solutions associated to the different families of admissible NEDs obtained from the combinations of asymptotic behaviours \textbf{B1}, \textbf{B2}, \textbf{B3} and \textbf{IRD} with the central-field behaviours \textbf{A1}, \textbf{A2} and \textbf{UVD}. Labels: \textbf{FEFS}($\equiv$ finite-energy flat-space solutions), \textbf{DEFS}($\equiv$ divergent-energy flat space solutions), \textbf{AS}($\equiv$ asymptotically Schwarzschild), \textbf{AA}($\equiv$ asymptotically anomalous), \textbf{NS}($\equiv$ naked singularities), \textbf{ EBH}($\equiv$ extreme black hole), \textbf{1hBH}($\equiv$ single-horizon black holes), \textbf{2hBH}($\equiv$ two-horizons black holes), \textbf{BP}($\equiv$ black points), \textbf{EBP}($\equiv$ extreme black points).}
  \label{table:I}
 \end{center}
\end{table}

\section{Thermodynamic analysis} \label{sectionIV}

Let us consider now the thermodynamic properties of the G-ESS black hole solutions of the admissible NEDs. In Ref.\cite{rasheed97} D. A. Rasheed established the validity of the zeroth and first laws of black hole thermodynamics for BH solutions of large classes of field theories minimally coupled to gravity, including the stationary solutions of G-NEDs with Maxwellian weak field limit. In the present context the rotational degrees of freedom are absent and the Rasheed proof is restricted to the non-rotating ESS black hole solutions, corresponding to our asymptotically coulombian cases \textbf{B2}. As we shall see, it is easy to extend the validity of this first law for non-rotating solutions including the cases \textbf{B1} and \textbf{B3} (the zeroth law is trivially satisfied for the ESS black hole solutions). However, as already mentioned, in the asymptotically \textbf{IRD} cases, for which the ADM mass is not defined and (as we shall see) the electrostatic potential diverges at infinity, the thermodynamic relations cannot be established, at least in the usual way. Consequently, the following considerations concern the \textbf{B} cases only.

\begin{figure}[h]
 \begin{center}
\includegraphics[width=0.75\textwidth]{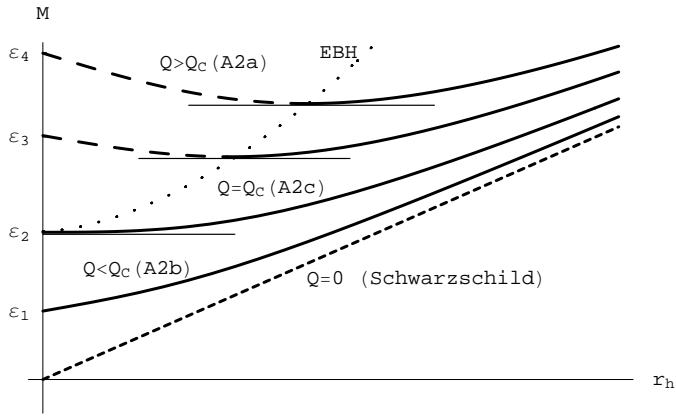}
\caption{$M-r_h$ diagram for the BH solutions of G-NEDs belonging to \textbf{A2} families for several values of $Q$. There are three subcases, corresponding to $Q \gtreqqless Q_{c} = (16 \pi a)^{-1}$. In case \textbf{A2a} the curves exhibit minima, corresponding to the EBH configurations. The dotted line linking these configurations, together with the short-dashed straight line of Schwarzschild BHs ($Q=0$), define the thermodynamically pertinent domain, corresponding to the external BH regions. The domain over the former curve corresponds to the BH interiors, where the dashed pieces of the constant-$Q$ lines give the $M-r_{h}$ relations for the inner horizons. The continuous pieces of the constant-$Q$ lines correspond to the outer (event) horizons, for which $M$ is always a monotonically increasing function of $r_{h}$ for the BH solutions of any admissible model. When $Q \rightarrow Q_{c}$ (case \textbf{A2c}) the EBH configurations converge towards a extreme BP, unique for a given \textbf{A2} NED. In the \textbf{A2b} case, for fixed $Q < Q_{c}$, we can have single-horizon BHs (with $M > \varepsilon_{1}$) and non-extreme BPs (with $M = \varepsilon_{1}$). This plot has been obtained from the BI model, as representative of the behaviour of the BHs belonging to the \textbf{A2} families (with $\sigma > 2$ in this case).}
\label{figure2}
 \end{center}
\end{figure}

\begin{figure}[h]
 \begin{center}
\includegraphics[width=0.75\textwidth]{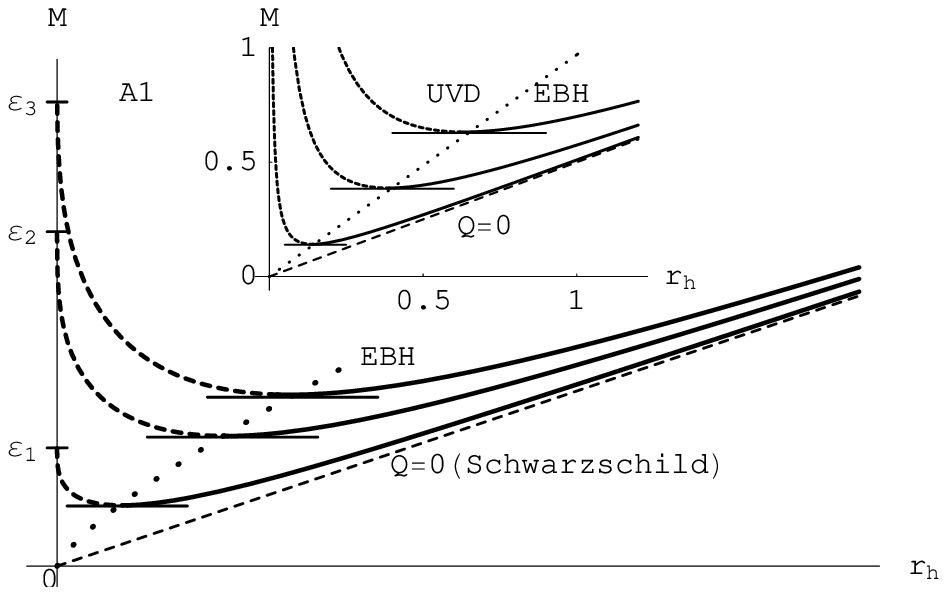}
\caption{$M-r_h$ curves at constant $Q$ for the \textbf{A1} (main frame) and \textbf{UVD} (small frame) families. The solid, dashed and dotted lines have the same meaning as in Fig.\ref{figure2}. The curves exhibit vertical asymptote at $r_{h} = 0$ in the \textbf{UVD} cases and are finite there, but with divergent slope, in the \textbf{A1} cases. For the \textbf{UVD} and \textbf{A1} cases we are plotting the behaviour of the BH solutions associated to the gravitating RN and EH models, respectively, as representative examples.}
\label{figure3}
 \end{center}
\end{figure}

The set of G-ESS (non-rotating) black hole solutions associated to a given G-NED can be assimilated to ``states" of a system characterized by two independent parameters, the most immediate ones being the integration constants: the mass $M$, well defined for asymptotically-\textbf{B} NEDs, and the charge $Q$. The location of the black hole event horizons is the largest solution  $r_{h}(M,Q)$ of the equation

\be
M = \frac{r_{h}}{2} + \varepsilon_{ex}(r_{h},Q),
\label{eq:(4-1)}
\en
obtained from the condition $g(r_{h},Q,M) = 0$ on Eq.(\ref{eq:(3-5)}). Owing to the admissibility requirements, $\varepsilon_{ex}(r,Q)$ is a monotonically decreasing and concave function of $r$ at constant $Q$, vanishing asymptotically as $r^{q+1}$ with $q<-1$ in the \textbf{B} cases \cite{dr09},\cite{dr10a}. It exhibits a vertical asymptote at $r=0$ in case \textbf{UVD} or takes, in cases \textbf{A1} and \textbf{A2}, a finite value there ($= Q^{3/2} \varepsilon(Q=1)$; see Eq.(\ref{eq:(2-16)bis})) with divergent slope in case \textbf{A1}. In case \textbf{A2} we distinguish three qualitatively different subcases, according to the values of the charge: case \textbf{A2a} if $Q > Q_{c}$, case \textbf{A2b} if $Q < Q_{c}$ and case \textbf{A2c} when $Q = Q_{c}$. The geometrical structures associated to these three subcases have been extensively discussed in Ref.\cite{dr10a}. The behaviour of $M$ as a function of $r_{h}$ for several representative values of $Q$, obtained from Eq.(\ref{eq:(4-1)}), is plotted in the figures \ref{figure2} and \ref{figure3} for the \textbf{A2} and \textbf{A1-UVD} cases, respectively. The locations of the minima of these curves (when present) correspond to the horizon radii ($r_{he}$) and masses ($M_{e}$) of extreme black holes, determined by equation (\ref{eq:(4-1)}) and the extremum condition

\be
\frac{\partial M}{\partial r_{h}}\Big \vert_{Q} = \frac{1}{2} -4\pi r_{h}^{2} T_{0}^{0}(r_{h},Q) = 0.
\label{eq:(4-1)bis}
\en
The cut points between the constant-mass horizontal lines and the positive-slope branches of the curves of constant charge, define the event horizon radii and the masses of the different black hole configurations. For a fixed $Q,$ if $M$ lies under the corresponding curve, there are no horizons and the solutions are naked singularities. Otherwise, besides extreme BHs, we have two-horizon (Cauchy and event) BHs, single horizon BHs (in cases \textbf{A1} and \textbf{A2}) and extreme (in case \textbf{A2c}) and non-extreme (in case \textbf{A2b}) black points. Moreover, since the radii of the external (event) horizons of the possible black hole configurations lie always on the positive-slope region of the constant-$Q$ curves, the mass-radius relation at fixed $Q$ is a monotonically increasing function for all the black hole solutions of the admissible G-NEDs. It approaches the Schwarzschild relation $M = r_{h}/2$ at large $r_{h}$, as well as at vanishing $Q$.

Let us define other important ``state functions" and establish the validity of the first law for the ESS black hole solutions associated to the set of all admissible NEDs of asymptotic types \textbf{B}.

From its general definition \cite{boyer69} the surface gravity ($\kappa$) for these charged BHs takes the form

\be
\kappa(r_{h},Q) = \frac{1}{2} \frac{\partial g(r,Q)}{\partial r}\Big \vert_{r=r_{h}} = \frac{1 - 8\pi r_{h}^{2} T_{0}^{0}(r_{h},Q)}{2r_{h}},
\label{eq:(4-2)}
\en
where Eqs.(\ref{eq:(2-13)}), (\ref{eq:(2-15)}) and (\ref{eq:(3-5)}) have been used. If we look now for the total differential of the ``energy" $M$ as a function of $r_{h}$ and $Q$ in Eq.(\ref{eq:(4-1)}), and using again Eqs.(\ref{eq:(2-13)}) and (\ref{eq:(2-15)}) we obtain for the partial derivatives the expressions

\be
\frac{\partial M}{\partial r_{h}}\Big \vert_{Q} = \frac{1 - 8\pi r_{h}^{2} T_{0}^{0}(r_{h},Q)}{2} = \kappa r_{h},
\label{eq:(4-3)}
\en
and

\be
\frac{\partial M}{\partial Q}\Big \vert_{r_{h}} = \frac{\partial \varepsilon_{ex}}{\partial Q}\Big \vert_{r_{h}} = 8\pi \int_{r_{h}}^{\infty} E(R,Q) dR = 8\pi A_{0}(r_{h},Q),
\label{eq:(4-4)}
\en
where $A_{0}(r_{h},Q)$ is the covariant time-like component (in the coordinate system (\ref{eq:(3-2)})) of the four-vector potential in the appropriate gauge, which is fixed through the conditions

\be
E(r,Q) = -\frac{d}{d r} A_{0}(r,Q) \hspace{.2cm};\hspace{.2cm} A_{0}(\infty,Q) = 0.
\label{eq:(4-5)}
\en
Thus, the total differential of $M$ reads

\be
dM = \kappa(r_{h},Q) r_{h} d r_{h} + 8\pi A_{0}(r_{h},Q) dQ.
\label{eq:(4-6)}
\en
With the usual definitions of the ``temperature" $T$ and the ``entropy" $S$ in terms of the surface gravity $\kappa$ and the horizon area $A,$ given by

\be
T = \frac{\kappa}{2\pi} = \frac{1 - 8\pi r_{h}^{2}T_{0}^{0}}{4\pi r_{h}}\hspace{.2cm}; \hspace{.2cm}
S = \frac{\textit{A}}{4} = \pi r_{h}^{2},
\label{eq:(4-7)}
\en
together with the usual notation

\be
8\pi A_{0}(r,Q) = \Phi(r,Q)
\label{eq:(4-7)bis}
\en
for the electric potential, the differential expression (\ref{eq:(4-6)}) takes the more familiar form

\bea
dM &=& T(r_{h},Q) dS + \Phi(r_{h},Q) dQ = \nonumber \\ &=& T(S,Q) dS + \Phi(S,Q) dQ.
\label{eq:(4-8)}
\ena
This is the expression of the first law, which holds for the black hole solutions of any G-NED of the nine families defined by the combinations of the three \textbf{B}-asymptotic behaviours and the three central-field behaviours \textbf{UVD}, \textbf{A1} and \textbf{A2}.

\subsection{The extreme black holes}

Let us now analyze the important case of the extreme BHs. From Eq.(\ref{eq:(4-1)bis}) and the first of (\ref{eq:(4-7)}) it is obvious that the temperature of these objects vanish. The condition (\ref{eq:(4-1)bis}) does not involve the mass and then, defines the radius of the extreme BH configurations as a function of the charge ($r_{he}(Q)$), which is monotonically increasing. Indeed, the general expression of the derivative of this function can be obtained from (\ref{eq:(4-1)bis}) and the first integral (\ref{eq:(2-4)}) in terms of the values of the field on the extreme-BH horizon. It reads

\be
\frac{d r_{he}}{d Q} = \frac{E_{e}}{r_{he} \varphi(E^{2}_{e})},
\label{eq:(4-1)ter}
\en
and is always positive for admissible models (for $Q > 0$ solutions). The explicit form of $r_{he}(Q)$ which, as we shall see at once, determines the whole thermodynamic behaviour of the subclass of extreme BHs, depends on the particular model and can be directly obtained from Eqs.(\ref{eq:(2-4)}), (\ref{eq:(2-13)}) and (\ref{eq:(4-1)bis}), once the form of the Lagrangian is specified. Alternatively, it can be obtained by integrating Eq.(\ref{eq:(4-1)ter}) with the appropriate boundary conditions. In fact this integration can be formally performed using the scale properties (\ref{eq:(2-4)bis}) leading, after some manipulations, to the general expression

\be
Q = Q_{0} \exp[-\Omega(R_{he})]\hspace{.2cm}; \hspace{.2cm} (R_{he} = \frac{r_{he}}{\sqrt{Q}})
\label{eq:(4-1)p}
\en
where the function $\Omega$ is defined by the integral

\be
\Omega(x) = 2\int_{x_{0}}^{x} \frac{z dz}{\left[z^{2} - \frac{2E(z,1)}{\varphi(E^{2}(z,1))}\right]},
\label{eq:(4-1)q}
\en
$Q_{0}$ and $x_{0}$ being related integration constants. For the Einstein-Maxwell theory this relation reduces to the simple and well known expression for Reissner-Nordstr\"om extreme BHs. But for general NEDs it can become largely involved. Nevertheless, the $r_{he}-Q$ relation for small and large radii is typical of each family. In both limits it can be determined by replacing the expression (\ref{eq:(2-8)}) (with $i=1,2$) in (\ref{eq:(4-1)bis}) and using the central or large-$r$ behaviours of the field ($E(r \rightarrow 0) \sim r^{p}$; $E(r \rightarrow \infty) \sim r^{q}$). This calculation leads to relations of the generic form

\be
r_{he} \sim \lambda_{i} Q^{\gamma_{i}},
\label{eq:(4-1)a}
\en
where $\gamma_{i}, (i=1,2)$ are the exponents defined in Eqs.(\ref{eq:(2-9)}) or (\ref{eq:(2-10)}) and the $\lambda_{i}$ are given by

\be
\lambda_{i} = \sqrt{\frac{\left[8\pi (2\gamma_{i} - 1)\right]^{(2\gamma_{i} - 1)}}{\alpha_{i} \gamma_{i}^{2\gamma_{i}}}},
\label{eq:(4-1)b}
\en
the $\alpha_{i}$ being the coefficients in Eq.(\ref{eq:(2-8)}).

In the \textbf{A2} cases the small-radius extreme BHs are present for charges slightly over the critical one ($Q \gtrsim Q_{c},$ see Fig.\ref{figure2}) and the charge-radius relation in this region can be obtained by a similar procedure from Eqs.(\ref{eq:(2-6)}) and (\ref{eq:(4-1)bis}), leading to

\be
r_{he} \sim \left[\frac{(2-\sigma)a}{2b_{0}} \frac{(Q - Q_{c})}{Q^{(1-\sigma/2)}} \right]^{1/\sigma},
\label{eq:(4-1)c}
\en
if $\sigma < 2$,

\be
r_{he} \sim \sqrt{\frac{2a}{\varphi(a^{2})} (Q - Q_{c})},
\label{eq:(4-1)d}
\en
if $\sigma > 2$ and

\be
r^{2}_{he} \ln(r_{he}) \sim \frac{a}{b_{0}} (Q_{c} - Q),
\label{eq:(4-1)e}
\en
when $\sigma = 2$. Once this $r_{he}-Q$ relation is obtained, Eq.(\ref{eq:(4-1)}) gives the mass of the extreme BHs as a function of their radius or their charge. The same will arise for the other state functions defined so far, fully determining the thermodynamic properties of the set of extreme BH configurations associated to any admissible NED. The lines of extreme BHs in the corresponding state diagrams are the $T=0$ isotherms which, as we shall see at once, define basic boundaries separating different phases in these diagrams and allow for the search of first-order phase transitions.

\subsection{The state variables and phase diagrams}

\begin{figure}[h]
 \begin{center}
\includegraphics[width=0.75\textwidth]{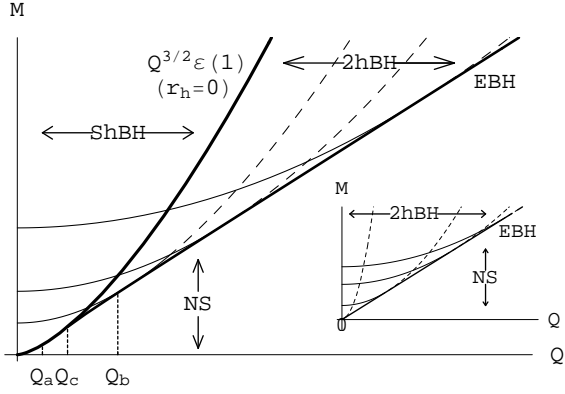}
\caption{Behaviour of $M$ as a function of $Q$ at fixed $r_{h}$ for the \textbf{A2} cases (main frame) represented here by the BI model and \textbf{UVD} cases (small frame) represented by the RN black hole. Some constant-$r_{h}$ curves are shown and are split (excepting for the $r_{h}=0$ one) into a continuous part, corresponding to the event horizons, and a dashed part, corresponding to the inner horizons. The EBH configurations are the envelopes of the beam of constant $r_{h}$ curves. The $r_{h} = 0$ curve in the \textbf{A2} cases (corresponding to $M = \varepsilon(Q) = Q^{3/2} \varepsilon(Q=1)$) defines the non-extreme BPs for charges below $Q_{c}$ and the EBP configuration for $Q = Q_c$. For $Q > Q_c$ this curve splits the regions associated to the single horizon and two-horizons BH configurations. The diagram corresponding to the \textbf{A1} cases is similar but with vanishing critical charge ($Q_{c} = 0$). The analysis of the diagram in the small frame (\textbf{UVD} cases) is similar, but now the $r_{h} \rightarrow 0$ curves approach the $Q = 0$ axis (which corresponds to the Schwarzschild BH limit) and the charged single-horizon BH configurations are absent.}
\label{figure9}
 \end{center}
\end{figure}

As already mentioned, the elementary solutions of a given gravitating NED are characterized by two ``state" parameters and it is possible to divide the plane of these parameters into separate regions corresponding to different phases of the set of solutions. The most immediate choice for this characterization is the pair of integration constants $M-Q$. It is possible to determine the qualitative form of the phase diagram for the different families of G-NEDs in terms of these parameters. To this end, let us analyze the behaviour of $M$ as a function of the charge at fixed $r_{h}$, which can be obtained from the scaling law \footnote{The following discussion on the relations between the different state variables (``equations of state") results from a (sometimes involved, but straightforward anyhow) analysis of their scale properties, their central and asymptotic behaviours and the admissibility conditions.}

\be
M(r_{h},Q) = \frac{(1-Q)\sqrt{Q}}{2} R_{h} + Q^{3/2} M(R_{h},Q=1)
\label{eq:(4-11)bis}
\en
($R_{h} = r_{h}/\sqrt{Q}$) obtained from (\ref{eq:(4-1)}) and the last of Eqs.(\ref{eq:(2-16)bis}). This function is plotted in figure \ref{figure9} for the cases \textbf{A1-A2} (main frame) and \textbf{UVD} (small frame). The gravitating BI and RN models, respectively have been taken as representative examples for this figure, but the phase diagrams for the models belonging to the corresponding families are qualitatively similar. The curves representing the extreme BH states are also displayed on both frames of this figure. They are the envelopes of the beams of constant $r_{h}$ curves (see Eq.(\ref{eq:(4-1)bis})) and begins tangent to the $r_{h}=0$ curve at $Q = Q_{c}$ in case \textbf{A2}. In case \textbf{A1} the corresponding curve is obtained by moving this tangency point to the origin ($Q_{c} \rightarrow 0,$ in the limit of large maximum field strength, $a \rightarrow \infty$). In cases \textbf{A1} and \textbf{A2}, the region over the $r_{h} = 0$ curve contains the single-horizon BH states, the region between this curve and the extreme BH curve contains the two-horizon BH states and the region under these curves and over the $Q$ axis corresponds to naked singularity configurations. Similar comments can be done for the \textbf{UVD} case (small frame), where the beam of constant-radius curves approach the $M$ axis as $r_{h} \rightarrow 0$.

As a function of $r_{h}$ at fixed $Q$ the behaviour of the mass has been already discussed at the beginning of this section and plotted in Figs. \ref{figure2} and \ref{figure3}.

\begin{figure}[h]
 \begin{center}
\includegraphics[width=0.75\textwidth]{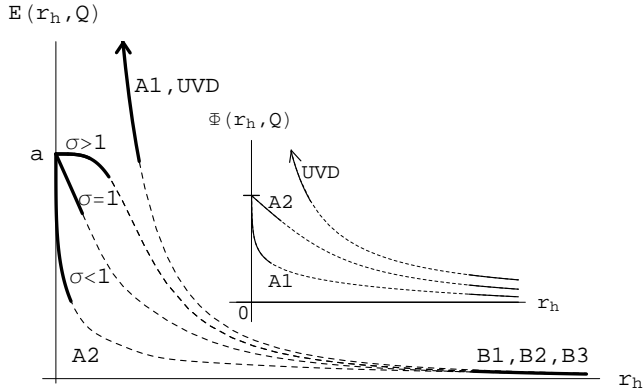}
\caption{Qualitative behaviour of $E(r_{h},Q)$ as a function of $r_{h}$ at fixed $Q$ (main frame). At the center, in cases \textbf{UVD} and \textbf{A1} the field diverges, while it takes a finite value in case \textbf{A2}. In this latter case the slope of $E(r,Q)$ at vanishing $r_{h}$ depends on the value of $\sigma$ (see Eq.(\ref{eq:(2-6)})). The small frame exhibits the corresponding behaviours of the potential $\Phi(r_h,Q)$. The dashed pieces of the lines, matching the central and asymptotic field behaviours, are always monotonically decreasing but can exhibit more involved forms than those showed here for many models (see the upper small frame in Fig.\ref{figure11} below). The thermodynamically pertinent parts of these curves start at the EBH radii $r_{he}(Q),$ obtained by solving Eq.(\ref{eq:(4-1)bis}) for the corresponding value of $Q$.}
\label{figure4}
 \end{center}
\end{figure}

Phase diagrams involving other state variables can be obtained by similar methods. Let us consider the case of the field on the horizon $E(r_{h},Q)$. As a function of $r_{h}$, at constant $Q$, its form is obtained from the first integral (\ref{eq:(2-4)}). The behaviour of this function around $r_{h} = 0$ and as $r_{h} \rightarrow \infty$ has been completely determined in section \ref{sectionII} for the elementary solutions of all classes of admissible NEDs and is plotted qualitatively in figure \ref{figure4}. It should be stressed that the only requirement to be satisfied by the field as function of $r$ at constant $Q$, in all admissible cases, is to be monotonically decreasing, but its second derivative can change sign many times at finite $r$, leading to less smooth lines than the dashed ones plotted in this figure (for an example, see the upper small frame in figure \ref{figure11} below). In determining the behaviour of $E(r_{h},Q)$ as a function of $Q$ at fixed $r_{h}$ we can use the scaling law (\ref{eq:(2-4)bis}), which leads to the condition

\begin{figure}[h]
 \begin{center}
\includegraphics[width=0.75\textwidth]{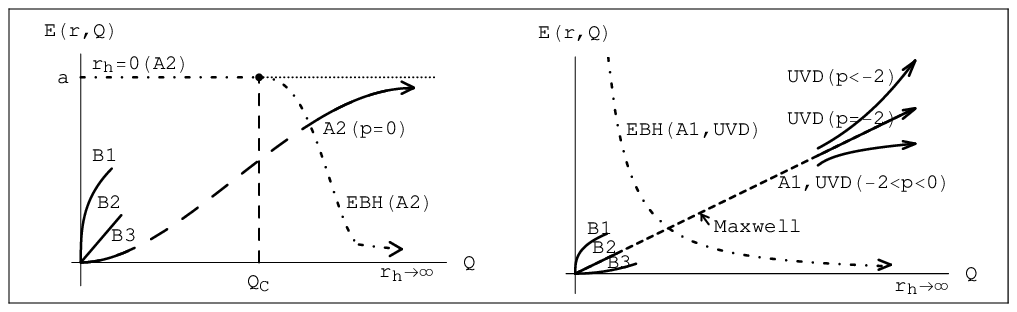}
\caption{Qualitative behaviour of $E(r_{h},Q)$ as a function of $Q$ at constant $r_{h}$, obtained from the scaling law (\ref{eq:(2-4)bis}) for the \textbf{A2-B} cases (left panel) and the \textbf{A1-B} and \textbf{UVD-B} cases. At small $Q$ the field vanishes with three different slopes, corresponding to the three possible large-$r$ behaviours. For large $Q$, in the \textbf{UVD} and \textbf{A1} cases (right panel) the field diverges as $Q \rightarrow \infty$, and exhibits different slopes depending on the degree of divergence at $r = 0$ ($p$ is the exponent in the first of Eqs.(\ref{eq:(2-5)})). In the \textbf{A2} cases (left panel) $E$ is limited from above at a fixed value ($E_{max}=a$) reached at $r_{h}=0$ for any $Q$. For $0 < Q < Q_{c}$ the dashed-dotted line $E = a$ corresponds to the set of non-extreme BPs. For $Q > Q_{c}$ the curve of EBHs is plotted here for \textbf{A2} cases with $\sigma \geq 3$. These curves define the upper boundary (dashed-dotted) of the thermodynamically pertinent region. For the \textbf{A1} and \textbf{UVD} cases this boundary is given by the corresponding curve of EBHs (dashed-dotted also). The points and branches over these boundaries correspond to the inner horizons. The matching between the small and large $Q$ parts of a constant-$r_{h}$ curve depends on the particular model, but is always monotonically increasing. The dashed piece of the straight line in the right panel represent this matching for the Maxwell (RN) case. The dashed piece of the curve in the left panel corresponds to the similar matching for a particular \textbf{A2-B3} model.}
\label{figure5}
 \end{center}
\end{figure}

\be
\frac{\partial E}{\partial Q}\Big \vert_{r_{h}} = -\frac{R_{h}}{2Q} \frac{d E(R_{h},Q=1)}{dR_{h}} > 0,
\label{eq:(4-9)}
\en
($R_{h} = r_{h}/\sqrt{Q}$) and we see that $E$, at constant $r_{h}$, is an increasing function of $Q$. Equations (\ref{eq:(2-4)bis}) and (\ref{eq:(4-9)}) allow to perform a complete analysis of the behaviour of $E$ around the limit $Q = 0$ ($R_{h} \rightarrow \infty$) and as $Q \rightarrow \infty$ ($R_{h} \rightarrow 0$) for the different families of NEDs considered here. This behaviour is plotted in figure \ref{figure5}.

For the extreme BHs we can obtain easily the behaviours of the field on the horizon at small and large $Q$ by replacing the $r_{he}-Q$ dependencies obtained for the different families. The large $Q$ behaviours and the small $Q$ behaviours in the cases \textbf{UVD} and \textbf{A1} are obtained from Eq.(\ref{eq:(4-1)a}) and have the same form, which reads

\be
E_{e}(Q) \sim \frac{\gamma_{i} Q^{-1}}{(8\pi)^{(2\gamma_{i}-1)}(2\gamma_{i}-1)},
\label{eq:(4-9)bis}
\en
as $Q \rightarrow 0$ ($i=1$) and as $Q \rightarrow \infty$ ($i=2$). In the same way, for the \textbf{A2} cases the behaviour of $E_{e}(Q)$ for small-radius extreme BHs ($Q \rightarrow Q_{c}$) is straightforwardly obtained from equations (\ref{eq:(4-1)b})-(\ref{eq:(4-1)e}) in models with different values of $\sigma$. For example, for $\sigma > 2$ we have

\be
E_{e}(Q) \sim a - b_{0} \left[\frac{2a}{\varphi(a^2)}\right]^{\sigma/2}  \left[\frac{Q - Q_{c}}{Q}\right]^{\sigma/2},
\label{eq:(4-9)ter}
\en
as $Q \rightarrow Q_{c}$.

\begin{figure}[h]
 \begin{center}
\includegraphics[width=0.75\textwidth]{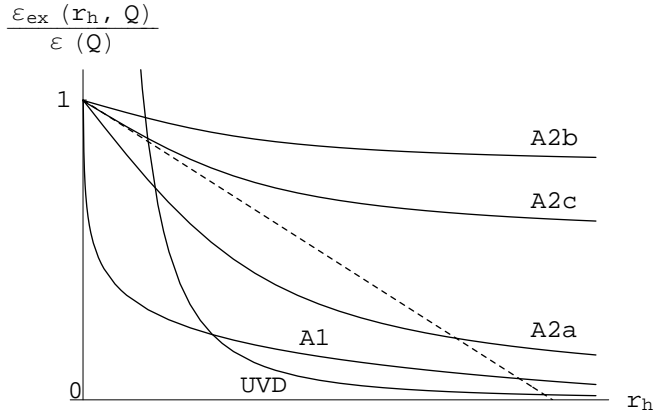}
\caption{Qualitative behaviour of $\varepsilon_{ex}(r,Q)$ as a function of $r_{h}$ at constant $Q$ for the \textbf{A1}, \textbf{A2} and \textbf{UVD} cases (in units of the total energy $\varepsilon(Q)$ in the \textbf{A1} and \textbf{A2} cases). The three \textbf{A2} curves correspond to three values of the charge: $Q > Q_{c}$ (cases \textbf{A2a}), $Q < Q_{c}$ (cases \textbf{A2b}) and  $Q = Q_{c}$ (case \textbf{A2c}). The cut points of the (dashed) straight lines defined by $y = 1 - r_{h}/(2\varepsilon(Q))$ with the \textbf{A2} curves give the location of the horizons (see Eq.(\ref{eq:(4-1)})).}
\label{figure7}
 \end{center}
\end{figure}

\begin{figure}[h]
 \begin{center}
\includegraphics[width=0.75\textwidth]{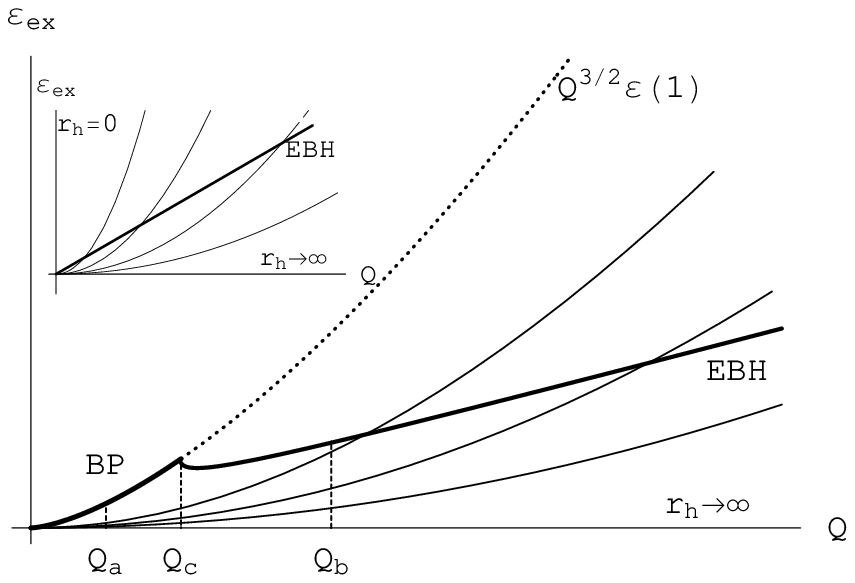}
\caption{Behaviour of $\varepsilon_{ex}(r_{h},Q)$ as a function of $Q$ at constant $r_{h}$ for the \textbf{A2} (main frame) and \textbf{UVD} (small frame) cases. The sets of EBH configurations are drawn in both cases. The beams of parabolic-like curves correspond to different constant values of $r_h$. The points of the region under the composite curve $r_{h} = 0$ and EBH in the case \textbf{A2}, and under the EBH curve in the \textbf{UVD} case correspond to the exterior of the BH configurations, whereas the points of the regions over these curves are associated to the BH interiors, playing no role in the thermodynamical analysis. The typical behaviours in the \textbf{A1} cases can be obtained from those of the main frame figure by allowing the cut point of the $r_{h} = 0$ and EBH curves to go to the origin, i.e. by making $Q_{c} \rightarrow 0$. In obtaining these curves we have used the BI and the RN models as representatives of the \textbf{A2} and \textbf{UVD} cases, respectively.}
\label{figure8}
 \end{center}
\end{figure}

Another important phase diagram can be obtained from the external energy function $\varepsilon_{ex}(r_{h},Q)$. As a function of $r_{h}$, for fixed $Q$, it has been extensively discussed and employed in Ref.\cite{dr10a} for the characterization of the different families. Its typical qualitative behaviour is plotted in figure \ref{figure7} for all cases. The dependence of $\varepsilon_{ex}$ on $Q$ at constant $r_{h}$ can be determined from the scaling law (\ref{eq:(2-16)bis}) and Eq.(\ref{eq:(4-4)}). This leads to the curves plotted in figure \ref{figure8}, corresponding to the cases \textbf{A1-A2} and \textbf{UVD} (small frame). The representative region for the event horizons of the BH configurations in the \textbf{A2} cases is bounded by above by the line of extreme BH configurations and the line $r_{h}= 0$, which intersect in a point with $Q = Q_{c}$, corresponding to the extreme black point configuration \textbf{A2c}. In case \textbf{A1} the line of extreme BHs lies always under the line $r_{h}= 0$ and defines the upper boundary of the BH region. This \textbf{A1} case is obtained from the main frame of Fig.\ref{figure8} in the limit of divergent maximum-field strength ($a \rightarrow \infty$ or $Q_{c} = 0$). For the \textbf{UVD} case (in the small frame) this boundary is also given by the line of extreme BHs. The $Q$ axis, corresponding to $r_{h} \rightarrow \infty$, is the lower boundary of this thermodynamically pertinent region.

\subsection{The potential $\Phi(r_{h},Q)$ and van der Waals-like transitions.} \label{sec:B}

\begin{figure}[h]
 \begin{center}
\includegraphics[width=0.75\textwidth]{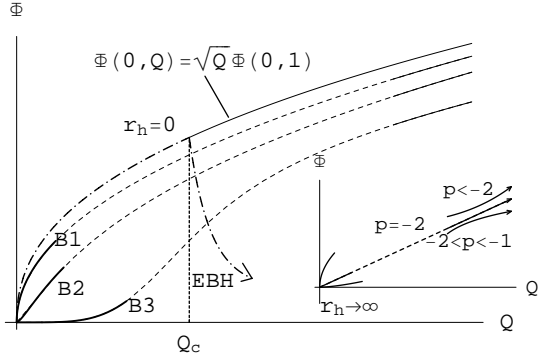}
\caption{Qualitative behaviour of $\Phi(r_{h},Q)$ as a function of $Q$ at constant $r_{h},$ for small and large values of $Q$, deduced from the scaling law (\ref{eq:(4-10)}) for the \textbf{A2} cases (main frame) and for the \textbf{UVD} and \textbf{A1} cases (small frame). As $Q \rightarrow 0$ the form of the constant-$r_{h}$ curves is governed by Eq.(\ref{eq:(4-10)d}), deduced from the asymptotic behaviour of $E(r_{h} \rightarrow \infty,Q)$ (\textbf{B}-cases). The upper curve in the main frame ($\Phi = \sqrt{Q} \Phi(0,1)$) gives the behaviour of the potential on the horizon of the BPs ($r_{h} = 0$) for $Q \leq Q_{c}$ in the case \textbf{A2}. The large-$Q$ shapes of the constant-$r_{h}$ curves are determined by the central behaviours of the field $E(r_{h} \rightarrow 0,Q)$. In cases \textbf{A2} these curves behave asymptotically as parabolic branches, parallel to the upper parabola (see Eq.(\ref{eq:(4-10)a})). The asymptotic shapes of the \textbf{A1} and \textbf{UVD} curves depend on the degree of divergence of the field $r=0$ (see Eqs.(\ref{eq:(4-10)b}) and (\ref{eq:(4-10)c})). The form of the dashed pieces of constant $r_{h}$ curves, matching the small and large-$Q$ regimes, depends on the particular model but must be always monotonically increasing. In fact, the thermodynamically meaningful region, where the external horizons are present, lies under the $r_{h} = 0$ curve for $Q < Q_{c}$ and the line of EBH solutions for $Q > Q_{c}$ in the \textbf{A2} cases. For the \textbf{A1} and \textbf{UVD} cases this region lies under the line of EBHs, which is calculated in the text for the different cases. It is qualitatively represented in the main frame for $Q \gtrsim Q_{c}$ in a particular \textbf{A2} model with $\sigma > 1$ (dashed-dotted line). The small-$Q$, $Q \gtrsim Q_{c}$ and large-$Q$ behaviours of these EBH curves defining those upper boundaries are plotted in Fig.\ref{figure6bis}) for all possible cases.}
\label{figure6}
 \end{center}
\end{figure}

As seen from the first law (\ref{eq:(4-8)}) the electrostatic potential is the conjugate state variable of the charge. Consequently the search for van der Waals-like phase transitions can be performed by analyzing the isotherms in the $\Phi-Q$ plane, in particular the set of extreme BHs which define the $T=0$ isotherm. In this way we need the expression of $\Phi$ as a function of $r_{h}$ and $Q$, which can be easily obtained from that of $E(r_{h},Q)$ through the relations (\ref{eq:(4-5)}). It is plotted at constant $Q$ in the small frame of figure \ref{figure4}. In looking for the behaviour of $\Phi$ as a function of $Q$ at constant $r_{h}$ we first obtain its scaling law from Eqs.(\ref{eq:(2-4)bis}) and (\ref{eq:(4-5)}), which reads

\be
\Phi(r_{h},Q) = \sqrt{Q} \Phi(R_{h},Q=1),
\label{eq:(4-10)}
\en
($R_{h} = r_{h}/\sqrt{Q}$) and shows that $\Phi$ is a homogeneous function of $r_{h}$ and $\sqrt{Q}$ of degree one. From this law we obtain the $Q$-derivative of $\Phi$ as

\be
\frac{\partial \Phi}{\partial Q}\Big \vert_{r_{h}} = \frac{\Phi(r_{h},Q) + 8\pi r_{h}E(r_{h},Q)}{2Q} > 0,
\label{eq:(4-11)}
\en
which shows that $\Phi$ is a monotonically increasing function of $Q$ at constant $r_{h}$. The form of this derivative and of the scaling law allows straightforwardly for a complete discussion of the behaviour of the potential at small and large $Q$ for constant $r_{h}$. This behaviour is displayed in figure \ref{figure6} for the different families. In the \textbf{A2} case, for $Q < Q_{c}$, the curve $r_{h} = 0$ corresponds to the non-extreme BPs and defines a part of the upper boundary of the thermodynamically pertinent region. The $Q = Q_{c}$ point of this curve defines the extreme BP. For large values of $Q$ the potential at constant $r_{h}$ behaves as

\be
\Phi(r_{h},Q \rightarrow \infty) \sim \sqrt{Q} \Phi(r_{h}=0,Q=1) - 8 \pi a r_{h}
\label{eq:(4-10)a}
\en
in the \textbf{A2} cases. In the \textbf{A1} cases we have

\be
\Phi(r_{h},Q \rightarrow \infty) \sim \sqrt{Q} \Phi(r_{h}=0,Q=1)
\label{eq:(4-10)b}
\en
whereas in the \textbf{UVD} cases we have

\be
\Phi(r_{h},Q \rightarrow \infty) \sim -8 \pi \frac{\nu_{1}(1)r_{h}^{p+1}}{p+1} Q^{-p/2},
\label{eq:(4-10)c}
\en
($p \leq -1$). At small $Q$ the behaviour of the constant $r_{h}$ curves is dominated by the large-$r$ behaviour of the electrostatic field, and reads

\be
\Phi(r_{h},Q \rightarrow 0) \sim -8 \pi \frac{\nu_{2}(1)r_{h}^{q+1}}{q+1} Q^{-q/2},
\label{eq:(4-10)d}
\en
where $-1 > q > -2$ (\textbf{B1}), $q = -2$ (\textbf{B2}) and $q < -2$ (\textbf{B3}).

For $Q > Q_{c}$ the thermodynamically pertinent region lies under the curve of extreme BHs, which is defined by eliminating $r_{he}$ between equation (\ref{eq:(4-1)bis}) and the expression of $\Phi(r_{h},Q)$. The shape of this curve, which begins on the extreme BP ($Q = Q_{c}$) in the cases \textbf{A2}, and at $Q = 0$ in cases \textbf{A1} and \textbf{UVD}, depends of the particular model, but its behaviour around these points and also for large values of $Q$ is universal for every family. It can be determined explicitly from equations (\ref{eq:(4-1)a}) to (\ref{eq:(4-1)e}) and the corresponding expansions of $\Phi(r,Q)$ around $r \rightarrow 0$ or $r \rightarrow \infty$, obtained from the integration of Eqs.(\ref{eq:(2-5)}) and (\ref{eq:(2-6)}). For large values of $Q$ these equations lead to

\begin{figure}[h]
 \begin{center}
\includegraphics[width=0.75\textwidth]{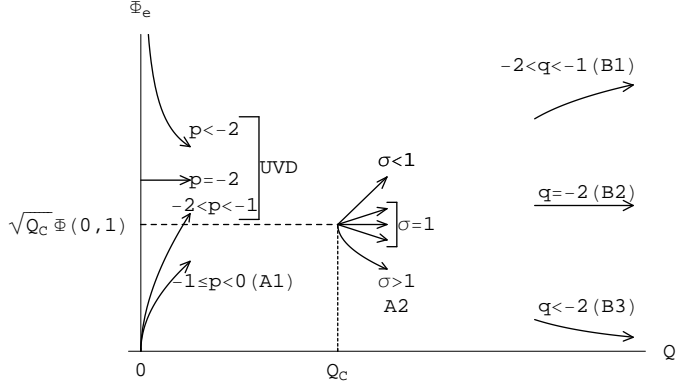}
\caption{Behaviours of the potential on the horizon of the extreme BHs ($\Phi_{e}(Q)$) for small values of $Q$ (in the \textbf{A1} and \textbf{UVD} cases), large values of $Q$ (in the \textbf{B} cases) and $Q \gtrsim Q_{c}$ (in the \textbf{A2} cases). See the discussion in the text for details.}
\label{figure6bis}
 \end{center}
\end{figure}

\be
\Phi_{e}(Q \rightarrow \infty) \sim \mu_{2} Q^{-\frac{q+2}{2q}},
\label{eq:(4-11)a}
\en
where the positive coefficient is

\be
\mu_{2} =  - \frac{2-q}{4(1+q)\sqrt{\alpha_{2}\gamma_{2}}} \left(\frac{2-q}{32\pi}\right)^{\frac{1}{q}}
\label{eq:(4-11)b}
\en
($q < -1$). We see that the potentials on the horizon for large-$Q$ extreme BH configurations diverge slower than $\sqrt{Q}$ in the \textbf{B1} cases, approach a constant value $\sqrt{8\pi/\alpha_{2}}$ (the same in all the asymptotically coulombian (\textbf{B2}) cases if the normalizing $\alpha_{2}$ constants of the Lagrangians in Eq.(\ref{eq:(2-8)}) are fixed to the same value) and vanish in the \textbf{B3} cases. For small $Q$ values in the \textbf{A1} cases the electrostatic potential on the horizon of extreme BHs behaves as

\be
\Phi_{e}(Q \rightarrow 0) \sim \Phi(0,1) \sqrt{Q}.
\label{eq:(4-11)c}
\en
In the \textbf{UVD} cases the same expressions (\ref{eq:(4-11)a}) and (\ref{eq:(4-11)b}) hold, with the replacements $q \rightarrow p$, $\alpha_{2} \rightarrow \alpha_{1}$, $\gamma_{2} \rightarrow \gamma_{1}$ and $\mu_{2} \rightarrow \mu_{1}$ (excepted in cases with $p = -1$, where logarithmic dependence arises, but does no modify the behaviours described here). Now, $\Phi_{e}(Q)$ vanishes at $Q = 0$ with divergent slope for $-2 < p \leq -1$ cases, takes a finite value there ($= \sqrt{8\pi/\alpha_{1}}$) if $p = -2$ and exhibits a vertical asymptote if $p < -2$ (see figure \ref{figure6bis}).

In the \textbf{A2} cases the integration of Eq.(\ref{eq:(2-6)}), together with (\ref{eq:(4-1)c})-(\ref{eq:(4-1)d}) lead, for $\sigma < 2$, to

\be
\Phi_{e}(Q \rightarrow Q_{c}) \sim \sqrt{Q}\left[\Phi(0,1) - \omega \left(1 - \frac{Q_{c}}{Q}\right)^{1/\sigma}\right],
\label{eq:(4-11)d}
\en
where

\be
\omega = 8\pi a^{1+1/\sigma} \left(\frac{2-\sigma}{2b_{0}}\right)^{1/\sigma},
\label{eq:(4-11)e}
\en
whereas for $\sigma > 2$ we obtain

\be
\Phi_{e}(Q \rightarrow Q_{c}) \sim \sqrt{Q}\left[\Phi(0,1) - 8\pi \sqrt{\frac{2a^{3}}{\varphi(a^{2})}} \sqrt{Q - Q_{c}} \right].
\label{eq:(4-11)f}
\en

The derivative of $\Phi_{e}(Q)$ can be directly obtained from Eq.(\ref{eq:(4-11)}) and (\ref{eq:(4-1)ter}) in the general case and reads

\be
\frac{d \Phi_{e}}{d Q} = \frac{1}{2Q} \left(\Phi_{e} - \frac{E_{e}}{r_{he} \varphi(E^{2}_{e})}\right).
\label{eq:(4-11)g}
\en
This equation allows the calculation of the slope of the extreme BH curves around $Q_{c}$, which is positive and finite in the cases $\sigma < 1$, finite positive, null or finite negative if $\sigma = 1$ and negative divergent if $\sigma > 1$. All these features are displayed in figure \ref{figure6bis}.

As already mentioned, the function $\Phi_{e}(Q)$ involves the two thermodynamically conjugate state variables $\Phi$ and $Q$, and defines a vanishing temperature ``isotherm" of the set of BH states associated to a given model (see equation (\ref{eq:(4-7)}) and the extremality condition (\ref{eq:(4-1)bis})). Thus the analysis of these curves can reveal the presence of van der Waals-like first-order phase transitions. This requires, as a necessary condition, the existence of some values of the charge for which the slope of $\Phi_{e}(Q)$ vanishes. Thus, the search for zeroes of (\ref{eq:(4-11)g}) is a first step in the identification of such transitions. A glance to figure \ref{figure6bis} shows that these special values of the charge must be necessarily present for models belonging to families with \textbf{A1} or \textbf{UVD} central behaviour (with $-2 < p < 0$) and \textbf{B3} asymptotic behaviour. The same arises for models belonging to the \textbf{UVD} (with $p < -2$) and \textbf{B1} families, or for those belonging to \textbf{A2-B1} families with $\sigma > 1$. Other interesting cases are the models belonging to the \textbf{A2-B2} family with $\sigma > 1$, when the condition

\be
\Phi(0,1) < 8\pi \sqrt{\frac{2a}{\alpha_{2}}}
\label{eq:(4-11)h}
\en
holds. But, beyond this necessary condition, the confirmation of the presence of such phase transitions in a given model requires the simultaneous analysis of the Gibbs free energy, defined as

\be
G = M - Q \Phi - TS = \frac{r_{h} - M}{2}
\label{eq:(4-11)i}
\en
for each set of extreme BHs. Similar methods have been applied to the search of this kind of phase transitions in the particular case of a Born-Infeld model in AdS background \cite{AR}. A more complete exploration of these issues is in progress.

\subsection{The temperature $T(r_{h},Q)$}

\begin{figure}[h]
 \begin{center}
\includegraphics[width=0.75\textwidth]{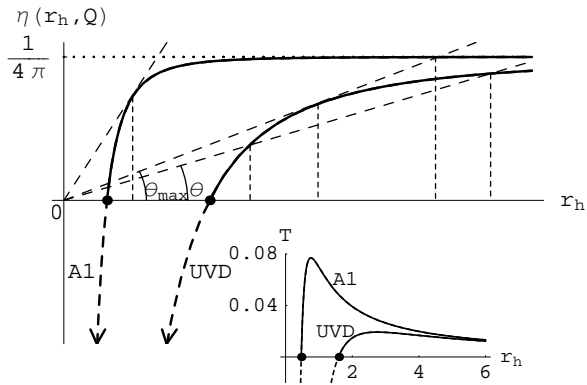}
\caption{Behaviour of $\eta(r_h,Q)$ as a function of $r_h$ and a fixed value of $Q$ for the \textbf{UVD} (RN) and \textbf{A1} (EH) cases. The slopes of the dashed straight lines matching the origin with the points of the $\eta$ curves give the temperatures of the associated BH configurations. The $\eta = 0$ points correspond to vanishing temperature EBH configurations. The associated temperature behaviours are plotted in the small frame.}
\label{figure10}
 \end{center}
\end{figure}

In analyzing the behaviour of the temperature (\ref{eq:(4-7)}) as a function of $r_{h}$ at constant $Q$ let us introduce the function

\be
\eta(r_{h},Q) = \frac{1}{2\pi} \frac{\partial M}{\partial r_{h}}\Big \vert_{Q} = \frac{1 - 8\pi r_{h}^{2}T_{0}^{0}(r_{h},Q)}{4\pi} = r_{h} T(r_{h},Q),
\label{eq:(4-12)}
\en
or, equivalently,

\be
\eta(r_{h},Q) = \frac{1}{4\pi} - 4Q E(r_{h},Q)\\
+ 2r_{h}^{2} \varphi(E^{2}(r_{h},Q)),
\label{eq:(4-12)bis}
\en
where the last hand side of Eq.(\ref{eq:(2-13)}) has been used. The derivative of this function is easily obtained and, as a consequence of the positivity of $\varphi$, satisfies

\be
\frac{\partial \eta}{\partial r_{h}}\Big \vert_{Q} = \frac{1}{2\pi}\frac{\partial^{2} M}{\partial r_{h}^{2}}\Big \vert_{Q} = 4r_{h}\varphi > 0
\label{eq:(4-13)}
\en
Thus $\eta$, at fixed $Q$, is a monotonically increasing function of $r_{h}$ for any admissible G-NED and, owing to the behaviour of $\varphi$ and $E$ as $r_{h} \rightarrow \infty$, approaches asymptotically the constant $1/4\pi$. The behaviour of this function is plotted in figure \ref{figure10} for representative examples of cases \textbf{A1} (Euler-Heisenberg) and \textbf{UVD} (Reissner-Nordstr\"om) for a given value of $Q$. The parts of these curves in the negative $\eta$ region correspond to the inner horizons and must be discarded for the thermodynamical analysis. The straight lines drawn from the origin to the points of these curves in the positive $\eta$ region have slopes $\tan(\theta) = T$, which define the corresponding temperatures. Since these curves are continuous, monotonically increasing and approach asymptotically the constant value $\eta = 1/4\pi$ for any $Q$, there is always an absolute maximum value of the angle $\theta$. Thus, the temperature ranges from zero, on the cut points with the $r_{h}$ axis (corresponding to the extreme BH configurations), to a maximum value $T_{max}$, corresponding to an angle $\theta_{max}$ in Fig.\ref{figure10}. This maximum (and, eventually, other local extrema of the curve) corresponds to BH configurations whose horizon radii are solutions of the equation

\begin{figure}[h]
 \begin{center}
\includegraphics[width=0.75\textwidth]{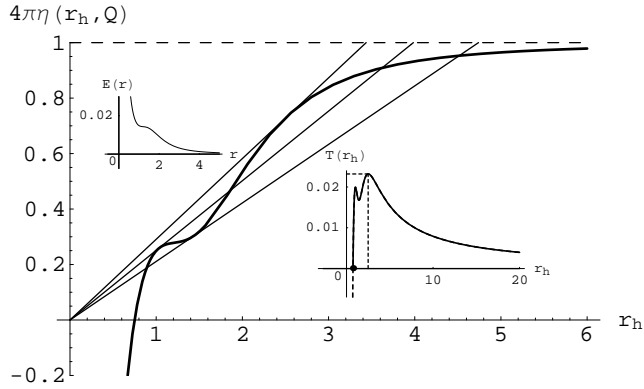}
\caption{Behaviour of the curve $\eta(r_h,Q)$ as a function of $r_h$ for BHs obtained from the gravitating field of Eq.(\ref{eq:(4-15)}), whose profile is shown in the small upper frame. Note that the change in the sign in the curvature of the field leads to the existence of up to three radial straight lines tangent to the curve $\eta(r_h,Q)$, which define an absolute maximum and two local extrema of the associated temperature curve (small lower frame).}
\label{figure11}
 \end{center}
\end{figure}

\be
\frac{\partial T}{\partial r_{h}}\Big \vert_{Q} = \frac{4r_{h}^{2} \varphi - \eta}{r_{h}^{2}} = 4\varphi - \frac{T}{r_{h}} = 0.
\label{eq:(4-14)}
\en
which results from Eqs.(\ref{eq:(4-12)})-(\ref{eq:(4-13)}). Moreover, from Eq.(\ref{eq:(4-12)}) and the asymptotic behaviour of the energy density in the \textbf{B}-cases it is easily seen that the temperature vanishes as

\be
T(r_{h} \rightarrow \infty,Q) \sim \frac{1}{4\pi r_{h}} \rightarrow 0,
\label{eq:(4-14)bis}
\en
for any large mass BH (this behaviour is related to the asymptotically Schwarzschild character of these space-times). The BH temperature as a function of $r_{h}$ at fixed $Q$ is plotted in the small frame of figure \ref{figure10}. The slope of this curve is given by the derivative defined in Eq.(\ref{eq:(4-14)}). The temperature as a function of $M$ has a similar shape, but the slope of the $T-M$ curve, obtained from Eqs.(\ref{eq:(4-12)}) and (\ref{eq:(4-14)}), reads

\be
\frac{\partial T}{\partial M}\Big \vert_{Q} = \frac{4r_{h}\varphi - T}{2\pi r_{h}^{2} T},
\label{eq:(4-14)ter}
\en
and diverges for the extreme BH configurations. Owing to the convex character of the curves $\eta(r_{h},Q)$ in the particular cases of figure \ref{figure10} there are at most two cut points with the radial lines for $T < T_{max}$ or, equivalently, there are at most two roots of Eq.(\ref{eq:(4-14)}) in these cases. However, in more general cases, the sign of the curvature of $\eta(r_{h},Q)$ at constant $Q$ can change many times as $r_{h}$ increases, leading to many cut points with the radial lines or, equivalently, to many solutions of equation (\ref{eq:(4-14)}). Let us illustrate this behaviour with a particular example. Consider a family of NEDs whose ESS solutions are of the form

\be
E(r,Q)= \frac{1}{R^{2}} + \frac{\alpha R^{n}}{\left(\beta + R^{n}\right)^{2}},
\label{eq:(4-15)}
\en
where $R = r/\sqrt{Q}$, the parameter $Q$ being the charge associated to a given solution and $\alpha > 0$, $\beta > 0$ and the integer $n \geq 2$ being fixed constants for a given model of the family. These fields are monotonically decreasing functions of $r$ if the parameters fulfill the (sufficient but not necessary) condition

\be
\alpha < \frac{6}{n+2} \beta^{\left(1 - \frac{1}{n}\right)},
\label{eq:(4-16)}
\en
and correspond to the ESS solutions of charge $Q$ associated to a family of admissible NEDs whose Lagrangian densities are parameterized by the constants $\alpha$, $\beta$ and $n$. The form of these Lagrangian densities can be directly obtained from Eqs.(\ref{eq:(2-4)}) and (\ref{eq:(4-15)}) through a quadrature. Owing to the central and asymptotic behaviour of the fields, the family belongs to the \textbf{UVD-B2} cases (as the Maxwell Lagrangian). The form of this field is plotted in the upper small frame of Fig.\ref{figure11} for the model with the particular choice of the parameters $n = 4$ and $\alpha=1/\beta=1/5.2$, satisfying the condition (\ref{eq:(4-16)}), and for the characteristic value of the charge $Q=1$. In the same figure we have plotted the corresponding function $\eta(r_{h},Q)$ and the associated temperature $T(r_{h},Q)$ (lower small frame). As can be seen there are in this example three solutions of Eq.(\ref{eq:(4-14)}) corresponding to a local maximum, a local minimum and an absolute maximum in the $T-r_{h}$ diagram. Because of the monotonic relation between $r_{h}$ and $M$, a similar behaviour must be obtained in a $T-M$ diagram, where the slope of the temperature decreases from $+\infty$ at the extreme BH configuration and vanishes on the first local maximum, defining a stable phase in this range. In the intermediate region between the local minimum and the absolute maximum, this slope becomes again positive, defining a second thermodynamically stable phase there. For increasing values of $n$ in the family (\ref{eq:(4-15)}) the number of solutions of Eq.(\ref{eq:(4-14)}) increases, and new ranges of $r_{h}$ between extrema support stable phases. We conclude that for models whose ESS black hole solutions belong to cases \textbf{UVD} and \textbf{A1} (with any \textbf{B}-type asymptotic behaviour) the temperature increases always from zero, for the extreme BH configurations, and vanish asymptotically for large mass BH configurations, as for the RN black holes. In the range of intermediate masses the temperature has always an absolute maximum and can exhibit a set of local extrema, determining a more or less involved phase structure, which is fully characterized by the solutions of Eq.(\ref{eq:(4-14)}).

\begin{figure}[h]
 \begin{center}
\includegraphics[width=0.75\textwidth]{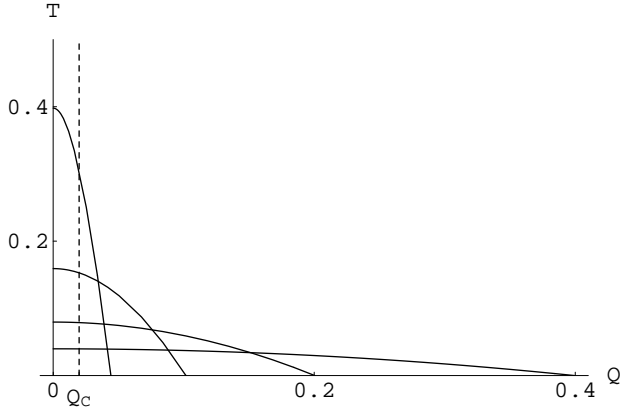}
\caption{Behaviour of $T(r_h,Q)$ as a function of $Q$ at constant $r_{h}$ for the Born-Infeld model, as representative of the \textbf{A2} cases. The set of EBH states lies on the $Q$ axis for $Q > Q_{c}$}.
\label{figure11b}
 \end{center}
\end{figure}

The behaviour of the temperature as a function of the charge at constant $r_{h}$ can be analyzed from its scaling law

\be
T(r_{h},Q) = \frac{1-Q}{4\pi \sqrt{Q}R_{h}} + \sqrt{Q} T(R_{h},Q=1),
\label{eq:(4-16)bis}
\en
($R_{h} = r_{h}/\sqrt{Q}$) which can be directly deduced from Eq.(\ref{eq:(4-12)bis}) and the first integral (\ref{eq:(2-4)}). From this equation we obtain the derivative

\be
\frac{\partial T}{\partial Q}\Big \vert_{r_{h}} = -\frac{4E(r_{h},Q)}{r_{h}},
\label{eq:(4-16)ter}
\en
which is negative for any $Q > 0$ and vanishes at $Q = 0,$ where the temperature reaches the Schwarzschild value $T = (4\pi r_{h})^{-1}$. Figure \ref{figure11b} displays the temperature as a function of the charge at different $r_{h}$ values for the case of field solutions $E(r)$ with monotonic slope (Born-Infeld here, as representative of \textbf{A2-B2} cases).

\subsection{The cases \textbf{A2} and the black points}

Although the validity of the analytic discussion of the precedent subsection is general, it has been applied there mostly to the thermal behaviour of BH solutions of \textbf{UVD} and \textbf{A1} families only. Let us discuss here the cases \textbf{A2}, which lead to more involved (and richer) thermodynamic structures. The method used in determining the behaviour of the temperature as a function of the horizon radius (or of the mass) at constant $Q$ still holds, but in this case the functions $\eta(r_{h},Q)$ exhibit new characteristics which lead to configurations with qualitatively new thermodynamic properties. The main difference with the previous cases concerns the behaviour of these functions as $r_{h} \rightarrow 0$, which defines the black point configurations. To address this question let us go to the next order in the polynomial expansion (\ref{eq:(2-6)}) of the field solutions of a given \textbf{A2} model around the center, which can be written as

\be
E(r) \sim a - b(Q) r^{\sigma} + c(Q) r^{\lambda} = a - b_{0} R^{\sigma} + c_{0} R^{\lambda},
\label{eq:(4-17)}
\en
($R = r/\sqrt{Q}$). Here $\lambda > \sigma$ and $Q^{\lambda/2} c(Q) = c(Q=1) = c_{0}$ is a universal constant of the model. From Eq.(\ref{eq:(4-12)bis}) we see that the behaviour of $\eta(r_{h} \sim 0,Q)$ can be obtained from Eq.(\ref{eq:(4-17)}) and the behaviour of $\varphi(r_{h},Q) \equiv \varphi(E^{2}(r_{h},Q))$ near the black point configurations ($r_{h} \sim 0$). Using (\ref{eq:(4-17)}) and the first integral (\ref{eq:(2-4)}) we are led to

\be
\varphi(r_{h} \sim 0,Q) \sim 2Q \left[\frac{\lambda c(Q)}{\lambda - 2} r_{h}^{\lambda - 2} - \frac{\sigma b(Q)}{\sigma - 2} r_{h}^{\sigma - 2}\right] + \Delta,
\label{eq:(4-18)}
\en
if $\sigma,\lambda \neq 2$. When $\lambda = 2$, $\varphi$ behaves as

\be
\varphi(r_{h} \sim 0,Q) \sim 2Q \left[2 c(Q) \ln(r_{h}) - \frac{\sigma b(Q)}{\sigma - 2} r_{h}^{\sigma - 2}\right] + \Delta,
\label{eq:(4-19)}
\en
and if $\sigma = 2$

\be
\varphi(r_{h} \sim 0,Q) \sim 2Q \left[\frac{\lambda c(Q)}{\lambda - 2} r_{h}^{\lambda-2} - 2b(Q) \ln(r_{h})\right] + \Delta.
\label{eq:(4-20)}
\en
In these equations the constants $\Delta$ are independent of the particular solution. If $\sigma > 2$ they are given by $\Delta = \varphi(a^{2}) > 0$. Otherwise they can be explicitly calculated from the form of the Lagrangians, as functions of the exponents $\sigma < 2,$ $\lambda$ and the associated universal coefficients in the polynomial expansion of the field around the center (this may require to go beyond the order of the expansion in Eq.(\ref{eq:(4-17)})). The corresponding expression for the \textbf{A2} cases with $\sigma < 2$ and $\lambda > 2$ is

\be
\Delta = \lim_{X \rightarrow a^{2}} \left[\varphi(X) - \frac{2 \sigma b_{0}^{2/\sigma}}{2 - \sigma} (a - \sqrt{X})^{\frac{\sigma-2}{\sigma}}\right].
\label{eq:(4-18)bis}
\en
while for $\sigma = 2$ we have

\be
\Delta = \lim_{X \rightarrow a^{2}} \left[\varphi(X) + 2b_{0} \ln(a - \sqrt{X}) \right].
\label{eq:(4-18)ter}
\en
The expression of $\Delta$ becomes more involved when $\lambda$ and some of the next-order exponents in the polynomial expansion of $E(r \sim 0)$ are smaller than two, but its calculation remains straightforward.

\begin{figure}[h]
 \begin{center}
\includegraphics[width=0.75\textwidth]{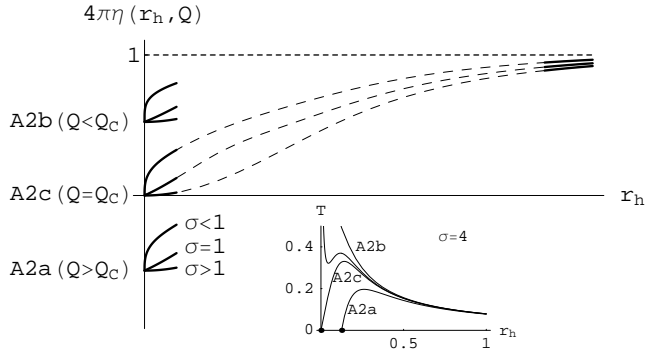}
\caption{Small and large $r_{h}$ behaviour of $\eta(r_h,Q)$ at constant $Q$ for the \textbf{A2} cases. The three beams formed by continuous pieces of curves at small $r_{h}$ correspond to the three subcases defined by the values of the charge $Q \gtreqless Q_c$. For each beam the three curves correspond to different values of the parameter $\sigma \gtreqless 1,$ exhibiting different slopes (vanishing, finite, or divergent, respectively). The asymptotic behaviour of these curves is obtained from Eqs.(\ref{eq:(4-12)bis}), (\ref{eq:(2-5)}) and (\ref{eq:(2-8)}) and is also displayed with pieces of continuous lines at large $r_{h}$. As in the previous cases (\textbf{A1} and \textbf{UVD}), the form of the dashed lines (intermediate region) depends on the particular model and can exhibit more involved shapes than the ones displayed in this plot. The small plot exhibits the form of the temperature for the BI model (for which $\sigma=4$) for several values of $Q$. Note that the temperature of the EBPs (cases \textbf{A2c}, with $Q = Q_{c}$) is given by the slope of the corresponding curve at $r_{h}=0$ and vanishes in models with $\sigma > 1$, takes a finite value in models with $\sigma = 1$ and diverges in models with $\sigma < 1$. The temperature of the non-extreme BPs (\textbf{A2b} cases, for which $Q < Q_{c}$) is always divergent.}
\label{figure12}
 \end{center}
\end{figure}

Equations (\ref{eq:(4-17)})-(\ref{eq:(4-18)ter}) replaced in (\ref{eq:(4-12)bis}) give the form of $\eta$ for the small radius black holes in the different \textbf{A2} cases. For the black points ($r_{h} = 0$), $\eta$ takes always the finite value

\be
\eta(r_{h} = 0,Q) = 4a(Q_{c} - Q)
\label{eq:(4-18)q}
\en
which vanishes for extreme BPs ($Q=Q_{c}$) and is positive for non-extreme BPs ($Q < Q_{c}$). If $Q > Q_{c}$, the small values of $r_{h} (< r_{he})$ correspond to inner horizons, for which $\eta(0,Q) < 0$ does not define a meaningful temperature (see Fig.\ref{figure12}). Obviously, for non-extreme BPs the temperature diverges in all cases, but the temperature of the extreme BPs is now dependent on the value of the parameter $\sigma$ of each \textbf{A2} model. Indeed, from the expansions of $\eta$ at small $r_{h}$ in the different cases we are led to the several possible expressions for the temperature of the small-radius BHs:

If $\sigma,\lambda \neq 2$ we have

\bea
T(r_{h} \rightarrow 0,Q) &\sim& \frac{\eta(r_{h} \rightarrow 0,Q)}{r_{h}} \sim \frac{1 - Q/Q_{c}}{4\pi r_{h}} - \frac{8Q b(Q)}{\sigma - 2} r_{h}^{\sigma - 1}\nonumber \\
&+& \frac{8Q c(Q)}{\lambda - 2} r_{h}^{\lambda - 1} + 2\Delta r_{h}.
\label{eq:(4-21)}
\ena
If $\sigma = 2$, we have

\bea
T(r_{h} \rightarrow 0,Q) &\sim& \frac{\eta(r_{h} \rightarrow 0,Q)}{r_{h}} \sim \frac{1 - Q/Q_{c}}{4\pi r_{h}} + \frac{8Q c(Q)}{\lambda - 2} r_{h}^{\lambda - 1} \nonumber \\
&+& 4Q b(Q) r_{h}(1 - 2 \ln(r_{h})) + 2\Delta r_{h},
\label{eq:(4-22)}
\ena
and if $\lambda = 2$,

\bea
T(r_{h} \rightarrow 0,Q) &\sim& \frac{\eta(r_{h} \rightarrow 0,Q)}{r_{h}} \sim \frac{1 - Q/Q_{c}}{4\pi r_{h}} - \frac{8Q b(Q)}{\sigma - 2} r_{h}^{\sigma - 1} \nonumber \\
&-& 4Q c(Q) r_{h}(1 - 2 \ln(r_{h})) + 2\Delta r_{h}.
\label{eq:(4-23)}
\ena
In all these expressions the first term dominates if $Q \neq Q_{c}$, in agreement with Eq.(\ref{eq:(4-18)q}). For the extreme BPs the temperature is given by the slope of $\eta(r_{h},Q_{c})$ in $r_{h}=0$ and is null if $\sigma > 1$, finite if $\sigma = 1$ or divergent if $\sigma < 1$ (see Fig.\ref{figure12}).

It is now evident that the temperature behaviour in case \textbf{A2a} ($\eta(0,Q > Q_{c}) < 0$) is similar to the ones of the cases \textbf{UVD} and \textbf{A1} (see the small frame of Fig.\ref{figure12}), exhibiting an extreme BH configuration with vanishing temperature at finite $r_{h}$, vanishing temperature BHs for large $r_{h}$, an absolute maximum temperature BH and (depending on the particular model) many possible BH configurations corresponding to local extrema of $T$ for some values of $r_{h}$ (or $M$), which define stable phases in some intermediate ranges. This behaviour is qualitatively independent of the values of the parameter $\sigma$.

In case \textbf{A2b} ($\eta(0,Q < Q_{c}) > 0$) there is a continuous spectrum of single-horizon BHs, whose radii increase from $r_{h} = 0$ (when $M_{BP} = \varepsilon(Q) = Q^{3/2} \varepsilon(Q = 1)$, corresponding to a non-extreme black point; see figure \ref{figure2}) to $r_{h} \rightarrow \infty$ as $M \rightarrow \infty$. In these cases the temperature diverges for the black points and vanishes asymptotically for large mass BHs. In the intermediate region, depending on the structure of the ESS fields, the temperature can exhibit local extrema defining ranges of stability, as in the example (\ref{eq:(4-15)}). Again, this behaviour is qualitatively independent on the values of the parameter $\sigma$.

In case \textbf{A2c} we have also a BH spectrum going from $r_{h} = 0$ (corresponding now to an extreme black point) to  $r_{h} \rightarrow \infty$. The mass of the extreme BPs equals the soliton energy and can be written as

\be
M_{BPextr} = Q_{c}^{3/2} \varepsilon(Q=1) = \frac{\Phi(r=0,Q=1)}{96(\pi a)^{3/2}},
\label{eq:(4-24)}
\en
where the last term is obtained from Eqs.(\ref{eq:(2-16)ter}), (\ref{eq:(4-5)}) and (\ref{eq:(4-7)bis}), and is a universal relation for \textbf{A2} models. The thermal behaviour of the extreme BPs for the different models is characterized by the values of the parameters $\sigma$ and $\lambda$ and can be directly obtained from Eqs.(\ref{eq:(4-21)})-(\ref{eq:(4-23)}) and their $r_{h}$ derivatives:

\begin{itemize}

\item For models with $\sigma < 1$ the temperature diverges for the extreme BPs, exhibiting a vertical asymptote at $r_{h} = 0$.

\item For models with $\sigma = 1$ the temperature of the extreme BP takes the finite value (see Eq.(\ref{eq:(4-21)}))

\be
T(r_{h} = 0,Q_{c}) = 8Q_{c}b(Q_{c}) = \frac{2b_{0}}{\sqrt{\pi a}},
\label{eq:(4-24)}
\en
which is a universal constant for each \textbf{A2} model. The slope of the temperature curve at vanishing $r_{h}$ diverges for $\lambda \leq 2$, being $\mp \infty$ for $c_{0} \gtrless 0$, respectively. For $\lambda > 2$ this slope is $2 \Delta$, which is finite. Here the sign of $\Delta$ determines the stability of the BPs.

\item For models with $1 < \sigma \leq 2$ the temperature of the BPs vanishes and its slope at $r_{h} = 0$ is always positively divergent.

\item For models with $\sigma > 2$ the black point temperatures vanish and the slope of the temperature curves at the origin is $2\Delta > 0$.

\end{itemize}

A similar analysis of the second derivatives of the temperature with respect to $r_{h}$, obtained from equations (\ref{eq:(4-21)})-(\ref{eq:(4-23)}), give us the concave or convex character of the $T-r_{h}$ curves for configurations around the BPs. A summary of these results for the extreme BP solutions in this \textbf{A2c} cases is given in table \ref{table:II}.

\begin{table}
 \begin{center}
   \begin{tabular}{| c | c | c | c| c |c |}
        \hline
      &\footnotesize $\sigma<1$ &\footnotesize $\sigma=1$  & \footnotesize $1<\sigma \leq 2$ & \footnotesize $\sigma>2$  \\
      \hline
      & & & & \\
      \footnotesize $T(r_{h}=0)$ & \footnotesize $+\infty$ & \footnotesize $\frac{2b_{0}}{\sqrt{\pi a}}$ & \footnotesize 0 & \footnotesize 0 \\
      & & & & \\
      \hline
       &  & \footnotesize $-\infty (c_{0}>0,\lambda\leq 2)$ & \footnotesize &
      \\ \footnotesize $\frac{\partial T}{\partial r_{h}}\vert_{r_{h}=0}$ & \footnotesize $-\infty$ & \footnotesize $+\infty (c_{0}<0,\lambda\leq 2)$  & \footnotesize $+\infty$ & \footnotesize $2\Delta$
      \\ & & \footnotesize $2\Delta (\lambda>2)$  &  & \\
      \hline
      &  & \footnotesize $c_{0} \cdot \infty (\lambda< 3)$ &  & \footnotesize $-\infty (2<\sigma <3)$
      \\ \footnotesize $\frac{\partial^2T}{\partial r_{h}^2}\vert_{r_{h}=0}$ & \footnotesize $+\infty$ & \footnotesize $64c_{0}\sqrt{\pi a}(\lambda=3)$  & \footnotesize $-\infty$ & \footnotesize $-64b_{0} \sqrt{\pi a}(\sigma=3)$
      \\ & & \footnotesize $0(\lambda>3)$  & & \footnotesize $0^- (\sigma>3)$ \\
      \hline
       &  & \footnotesize $0^-(\lambda< 3,\Delta=0)$ & &
      \\ $ \footnotesize C_{Q,r_{h}=0}$ & \footnotesize $0^{-}$ & \footnotesize $\frac{b_{0}}{16 a c_{0}}(\lambda=3, \Delta=0)$& \footnotesize $0^+$& \footnotesize $0^+$
      \\ & & \footnotesize $+\infty(\lambda>3,\Delta=0)$  & & \\
      & & \footnotesize $0 \cdot \Delta $ & & \\
      \hline
   \end{tabular}
   \caption{Thermodynamic behaviour for extreme BPs ($r_h=0$ and $Q=Q_{c}$). See Eqs.(\ref{eq:(4-17)})-(\ref{eq:(4-18)ter}) for the meanings of the parameters.}
 \label{table:II}
 \end{center}
\end{table}

\subsection{The specific heats $C_{Q}(r_{h},Q)$}

The specific heat at constant charge of the BH configurations is defined as

\be
C_{Q}(r_{h},Q) = \frac{\partial M}{\partial T}\Big \vert_{Q} = \frac{\frac{\partial M}{\partial r_{h}}\Big \vert_{Q}}{\frac{\partial T}{\partial r_{h}}\Big \vert_{Q}} = \frac{2\pi \eta}{\frac{\partial T}{\partial r_{h}}\mid_{Q}},
\label{eq:(4-25)}
\en
and we see that, as a function of $r_{h}$ or $M$, diverges on the relative extrema of the temperature function, exhibiting there vertical asymptote. From Eqs.(\ref{eq:(4-12)}) and (\ref{eq:(4-12)bis}) we obtain the general formula

\be
C_{Q}(r_{h},Q) = \frac{2\pi r_{h}^{2} T}{4r_{h} \varphi - T},
\label{eq:(4-26)}
\en
and eliminating the Lagrangian function $\varphi$ in terms of the temperature and the field we are led to

\be
C_{Q}(r_{h},Q) = \frac{4 \pi^{2} r_{h}^{3} T}{2\pi r_{h} T + 16\pi Q E - 1}.
\label{eq:(4-26)bis}
\en
We see from this last expression and the asymptotic behaviour of $T$, given by Eq.(\ref{eq:(4-14)bis}), that the specific heat of large mass BH solutions of any admissible \textbf{B}-type model behaves as

\be
C_{Q}(r_{h} \rightarrow \infty,Q) \sim -2\pi r_{h}^{2} \rightarrow -\infty,
\label{eq:(4-27)}
\en
as should be expected from their asymptotically Schwarzschild character. The slope of $C_{Q}$ as a function of $r_{h}$ reads

\be
\frac{\partial C_{Q}}{\partial r_{h}}\Big \vert_{Q} = 2\pi r_{h} + \frac{4\pi r_{h} T}{4r_{h} \varphi - T} \left[1 - \frac{4Q \frac{dE}{dr_h} + T/2}{4r_{h} \varphi - T}\right].
\label{eq:(4-27)bis}
\en
Using the asymptotic behaviours of the different functions in this formula for the \textbf{B}-cases we see that for large-$r_{h}$ BHs this derivative diverges as $-4\pi r_{h}$, consistently with the asymptotic behaviour of $C_{Q}$ in Eq.(\ref{eq:(4-27)}).

For the extreme BHs associated to the cases \textbf{UVD}, \textbf{A1} and \textbf{A2a} we have $\eta = T = 0$ and $\frac{\partial T}{\partial r_{h}}\big \vert_{Q} = 4\varphi > 0$ (see Eq.(\ref{eq:(4-13)})). Consequently $C_{Q;extr} = 0$ for any extreme BH configuration of admissible models. The slope of the specific heat as a function of $r_{h}$, given by Eq.(\ref{eq:(4-27)bis}), converges to the value

\be
\frac{\partial C_{Q}}{\partial r_{h}}\Big \vert_{Q;extr} = 2\pi r_{he},
\label{eq:(4-27)ter}
\en
for the extreme BHs of any admissible model. The slope of the specific heat as a function of the mass is obtained from Eq.(\ref{eq:(4-27)bis}) as

\be
\frac{\partial C_{Q}}{\partial M}\Big\vert_{Q} = \frac{\frac{\partial C_{Q}}{\partial r_{h}}\Big\vert_{Q}}{\frac{\partial M}{\partial r_{h}}\Big\vert_{Q}} = \frac{1}{2\pi r_{h} T} \frac{\partial C_{Q}}{\partial r_{h}}\Big\vert_{Q},
\label{eq:(4-27)cuart}
\en
and diverges in the extreme limit as $1/T$.

Let us analyze now the specific heats of the BP configurations discussed above. Using Eq.(\ref{eq:(4-26)bis}) and replacing there the expressions (\ref{eq:(4-21)}) and (\ref{eq:(4-17)}) for the expansion of the temperature and the field around $r_{h} \sim 0$ we are led to

\be
C_{Q}(r_{h} \sim 0,Q) \sim  \frac{N}{D}
\label{eq:(4-28)}
\en
where

\bea
N & = & 2\pi \left(1-\frac{Q}{Q_{c}}\right)r_{h}^{2} + \frac{64 \pi^{2} Q b(Q)}{2-\sigma} r_{h}^{\sigma+2} + \nonumber \\ &+& \frac{64 \pi^{2} Q c(Q)}{\lambda-2}r_{h}^{\lambda+2} + 16 \pi^{2}\Delta r_{h}^{4}
\label{eq:(4-28)bis}
\ena
and

\bea
D & = & \left(\frac{Q}{Q_{c}} - 1\right) - \frac{32\pi(\sigma-1) Q b(Q)}{\sigma-2} r_{h}^{\sigma} + \nonumber \\ &+&\frac{32\pi Q (\lambda-1) c(Q)}{\lambda-2} r_{h}^{\lambda} + 8\pi \Delta r_{h}^{2},
\label{eq:(4-28)ter}
\ena
which is valid for $\sigma,\lambda \neq 2$. It is straightforward to obtain two similar expressions for the cases $\sigma = 2$ and $\lambda = 2$ from Eqs.(\ref{eq:(4-22)}) and (\ref{eq:(4-23)}). The analysis of these equations leads immediately to the following conclusions:

\begin{itemize}

\item The specific heats vanish for any non-extreme BP (\textbf{A2b} cases, $Q < Q_{c}$), behaving as

\be
C_{Q}(r_{h} \rightarrow 0,Q) \sim -2\pi r_{h}^{2} \rightarrow 0^{-},
\label{eq:(4-29)}
\en
for small horizon radius BHs.

\item For extreme BPs (\textbf{A2c} cases, $Q = Q_{c}$) in models with $\sigma > 2$ the specific heats vanish, behaving as

\be
C_{Q}(r_{h} \rightarrow 0,Q_{c}) \sim 2\pi r_{h}^{2} \rightarrow 0^{+},
\label{eq:(4-30)}
\en
for small horizon radius BHs.

\item For extreme BPs in models with $1 < \sigma \leq 2$ the specific heats vanish, behaving as

\be
C_{Q}(r_{h} \rightarrow 0,Q_{c}) \sim \frac{2\pi}{\sigma - 1} r_{h}^{2} \rightarrow 0^{+},
\label{eq:(4-31)}
\en
for small horizon radius BHs.

\item For extreme BPs in models with $\sigma < 1$ the specific heats vanish, behaving as

\be
C_{Q}(r_{h} \rightarrow 0,Q_{c}) \sim \frac{2\pi}{\sigma - 1} r_{h}^{2} \rightarrow 0^{-},
\label{eq:(4-32)}
\en
for small horizon radius BHs.

\item For extreme BPs in models with $\sigma = 1$ the dominant terms in the expression of the specific heat of small radius BH configurations reduce to

\be
C_{Q}(r_{h} \rightarrow 0,Q_{c}) \sim \frac{2\pi b(Q_{c}) (\lambda - 2)r_{h}}{c(Q_{c}) (\lambda - 1) r_{h}^{\lambda-2} + 4\pi a(\lambda - 2)\Delta},
\label{eq:(4-33)}
\en
when $\lambda \neq 2$ and

\be
C_{Q}(r_{h} \rightarrow 0,Q_{c}) \sim \frac{2\pi b(Q_{c}) r_{h}}{c(Q_{c})\ln(r_{h})} = \sqrt{\frac{\pi}{a}} \frac{b_{0} r_{h}}{2c_{0} \ln(r_{h})} \rightarrow 0,
\label{eq:(4-34)}
\en
when $\lambda = 2$. As easily seen, the specific heats of extreme BPs in these $\sigma = 1$ cases vanish, excepted for models with $\lambda \geq 3$ and $\Delta = 0$, as we shall see at once. If $\lambda = 2$, $C_{Q}$ behaves for small horizon radius BHs as in Eq.(\ref{eq:(4-34)}), where the sign near zero is given by that of the coefficient $c_{0}$. This behaviour is not dependent on the value of $\Delta$. For $\lambda > 2$ and $\Delta \neq 0$, the specific heat of small-$r_{h}$ BHs behaves as

\be
C_{Q}(r_{h} \rightarrow 0,Q_{c}) \sim \sqrt{\frac{\pi}{a}} \frac{2b_{0} r_{h}}{\Delta} \rightarrow 0,
\label{eq:(4-35)}
\en
where $\Delta$ determines now the sign of the approach to zero. For $1 < \lambda < 2$ the specific heat behaves as

\be
C_{Q}(r_{h} \rightarrow 0,Q_{c}) \sim \frac{2\pi b_{0} (\lambda - 2)}{c_{0}(\lambda - 1)} (16 \pi a)^{\frac{\lambda-1}{2} r_{h}^{3 - \lambda} } \rightarrow 0,
\label{eq:(4-36)}
\en
which vanishes with the opposite sign of $c_{0}$ and is not dependent on the value of $\Delta$. Finally, for models with $\Delta = 0$ the expression (\ref{eq:(4-36)}) of $C_{Q}$ remains valid for any value of $\lambda \neq 2$. In these cases we can have extreme BP solutions with finite specific heat

\be
C_{Q}(r_{h} = 0,Q_{c}) = \frac{b_{0}}{16 a c_{0}},
\label{eq:(4-37)}
\en
if $\lambda = 3$, besides models with divergent specific heat extreme BPs if $\lambda > 3$.

\end{itemize}

In the last line of table \ref{table:II} we have included these results on the possible behaviours of the specific heats for extreme black point solutions associated to the family of \textbf{A2}-type gravitating NEDs. This discussion exhausts the possible thermodynamic structures of the G-ESS black hole solutions of admissible NEDs minimally coupled to gravity.

\subsection{Some examples}

In order to illustrate different points of this analysis let us consider some particular examples belonging to the different families of admissible models.

\begin{figure}[h]
 \begin{center}
\includegraphics[width=0.75\textwidth]{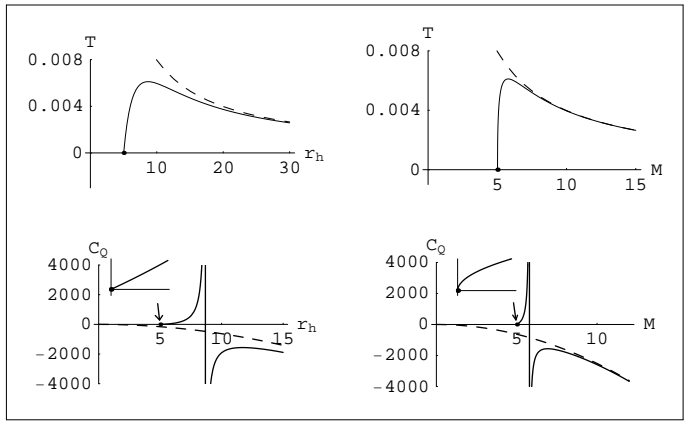}
\caption{Behaviour of the temperature and specific heat as functions of $r_h$ and $M$ at constant $Q$ for the RN BH solutions. The specific heats exhibit vertical asymptote on the maxima of the temperature curves. The slope of the temperature as a function of $M$ diverges for EBH configurations. The small frames expand the detail of the $C_{Q}$ curves around these EBH solutions (see Eqs.(\ref{eq:(4-27)bis}) and (\ref{eq:(4-27)cuart})). The dashed lines show the same functions for the Schwarzschild BHs.}
\label{figure13}
 \end{center}
\end{figure}

The thermodynamic properties of the Reissner-Nordstr\"om solution of the Einstein-Maxwell field equations are representative of the typical behaviour of the small-radius BH solutions of the \textbf{UVD} family. Although the gravitating counterpart of this NED has been extensively analyzed in the literature (see \cite{RN-thermo} and references therein) let us mention a few of its features, owing to this representative character. We have already make use of this example in the small frames of figures \ref{figure3} and \ref{figure10} representing the generic behaviour of the mass and temperature as functions of the BH radius for this kind of solutions. We have also displayed the temperature and the specific heat as functions of the mass and of the horizon radius in figure \ref{figure13}. The explicit expressions for these variables in the RN case are straightforwardly obtained from the formulae of the precedent analysis and lead to the expected thermodynamical behaviour for the small and large horizon radius BHs of the \textbf{UVD-B2} cases.

As a representative example of the \textbf{A1} family let us consider the Euler-Heisenberg effective model of quantum electrodynamics \cite{EH36,dobado97}, whose gravitating counterpart has been analyzed in Ref.\cite{Yajima01}. As discussed in the introduction, this is an effective model accounting (to leading order) for screening effects of the Dirac vacuum on electromagnetic wave propagation. The interest on this model in the present context is related to the possibility of explore modifications induced by quantum corrections of the electrostatic field on the gravitational field generated by point-like charges (which, in absence of these corrections is described by the Reissner-Nordstr\"om BH solutions). The Lagrangian in this case reads \cite{EH36}

\be
\varphi(X,Y) = \frac{X}{2} + \mu \left(X^{2} + \frac{7}{4}Y^{2}\right),
\label{eq:(4-38)}
\en
$\mu$ being a positive constant. The ESS field solution $E(R)$ is the unique positive root of the equation

\be
2\mu E^{3} + \frac{E}{2} - \frac{1}{R^{2}} = 0.
\label{eq:(4-39)}
\en
($R = r/\sqrt{Q}$). As easily seen, this field behaves at the center and asymptotically as

\bea
E(R \rightarrow 0) &\sim& 2\mu^{-1/3} R^{-2/3} \rightarrow \infty \nonumber \\ E(R \rightarrow \infty) &\sim& R^{-2} \rightarrow 0,
\label{eq:(4-40)}
\ena
and the model belongs to the \textbf{A1-B2} family, whose ESS solutions in flat space are of finite energy \cite{dr09}. The temperature of the BH solutions of this model, as a function of $r_{h}$, can be obtained from Eq.(\ref{eq:(4-12)bis}) in terms of the root of (\ref{eq:(4-39)}) as

\be
T(r_h,Q) = \frac{1}{4\pi r_{h}} - 3\sqrt{Q} \frac{E}{r_{h}} + \frac{r_{h}}{2\sqrt{Q}} E^{2},
\label{eq:(4-41)}
\en
and the expression for the specific heat results directly from Eq.(\ref{eq:(4-26)bis}).

The mass as a function of $r_{h}$ for several values of $Q$ has been plotted in figure \ref{figure3} for this EH model (main frame), where also the behaviour of the RN model is displayed (small frame). The only qualitative difference between both models arises on the small-$r_{h}$ region, where the EH model leads to one-horizon BH solutions for masses over the total electrostatic energy ($M > \varepsilon(Q)$). In the thermodynamically pertinent regions in these figures, which lie between the (dotted) lines of extreme BHs and the (dashed) line of Schwarzschild BHs, the behaviour of the BH masses versus $r_{h}$ are qualitatively the same. Moreover, the behaviours of the functions $\eta(r_{h})$ and $T(r_{h})$ at fixed $Q$ for these models (plotted in figure \ref{figure10}) as well as the behaviour of the specific heats (plotted in figure \ref{figure13} for the RN model) are also similar. These results illustrate the qualitative similarity of the thermodynamic properties of the BH solutions of \textbf{UVD} and \textbf{A1} families near the extreme BH configurations.

A first representative example of the \textbf{A2} family is the well known Born-Infeld model \cite{BI34}. As mentioned in the introduction, this model is a nonlinear generalization of classical electrodynamics which avoids the problem of the divergent self-energy of point-like charge solutions. Moreover, BI-like models arise as effective gauge field theories in the low-energy regime of string theory \cite{ST}. Besides their use in the study of black hole physics \cite{BI-AdS,BI-AdS-higher} they have found applications in cosmological contexts \cite{darkenerg}. The corresponding Lagrangian density reads

\be
\varphi(X,Y) = \frac{1 - \sqrt{1 - \mu^{2}X - \frac{\mu^{4}}{4} Y^{2}}}{\mu^{2}},
\label{eq:(4-42)}
\en
where $1/\mu$ is the maximum field strength. When $\mu \rightarrow 0$ in Eq.(\ref{eq:(4-42)}), the Maxwell Lagrangian ($\varphi=X/2$) is recovered. The ESS solutions corresponding to model (\ref{eq:(4-42)}) take the form

\be
E(R,Q) = \frac{2}{\sqrt{R^{4} + 4\mu^{2}}}; \hspace{.2cm} (R = r/\sqrt{Q}).
\label{eq:(4-43)}
\en
The expansion of these solutions around $r = 0$

\be
E(R,Q) = \frac{1}{\mu} - \frac{R^{4}}{8\mu^{3}} + \frac{3R^{8}}{128\mu^{5}}; \hspace{.2cm} (R = r/\sqrt{Q}).
\label{eq:(4-44)}
\en
gives the values of the maximum field strength $a = 1/\mu$, the coefficients $b_{0} = a^{3}/8$, $c_{0} = 3a^{5}/128$ and the exponents $\sigma = 4$ and $\lambda = 8$ (see Eq.(\ref{eq:(4-17)})). The constant $\Delta$ in the expansion (\ref{eq:(4-18)}) of the Lagrangian density is now $\Delta = 1/\mu^{2} = a^{2}$.

The expression of the BH temperature as a function of $r_{h}$ and $Q$ is obtained from the equations (\ref{eq:(4-12)bis}), (\ref{eq:(4-42)}) and (\ref{eq:(4-43)}) and reads

\be
T(r_{h},Q) = \frac{1}{4\pi r_{h}} + \frac{2r_{h}}{\mu^{2}}\left(1 - \sqrt{1 + \frac{4\mu^{2} Q^{2}}{r_{h}^{4}}}\right),
\label{eq:(4-45)}
\en
which behaves as

\be
T(r_{h} \rightarrow 0,Q) \sim  \frac{1 - Q/Q_{c}}{4\pi r_{h}},
\label{eq:(4-46)}
\en
for small radius BHs. So we recover the results of the precedent analysis on the temperature of the BH configurations close to the extreme ($Q = Q_{c}$) and non-extreme ($Q < Q_{c}$) BPs, described by equation (\ref{eq:(4-21)}). For values of the charge $Q > Q_{c}$ (\textbf{A2a} cases) the radii of the extreme BH configurations as functions of $Q$, obtained from Eq.(\ref{eq:(4-1)bis}), take the form

\be
r_{he}(Q > Q_{c}) = 4\pi Q \sqrt{1 - \left(\frac{Q_{c}}{Q}\right)^{2}}
\label{eq:(4-46)bis}
\en
and, as expected, the temperature (\ref{eq:(4-45)}) vanishes for these configurations.

Similar considerations can be done for the behaviour of the specific heat, which results directly from Eqs.(\ref{eq:(4-26)bis}), (\ref{eq:(4-43)}) and (\ref{eq:(4-45)}). For small radius BHs the behaviour of $C_{Q}$ is obtained from these formulae or, alternatively, from Eq.(\ref{eq:(4-29)}) in case \textbf{A2b} or from Eq.(\ref{eq:(4-30)}) in case \textbf{A2c}, once the characteristic parameters of the model have been fixed.

\begin{figure}[h]
 \begin{center}
\includegraphics[width=0.75\textwidth]{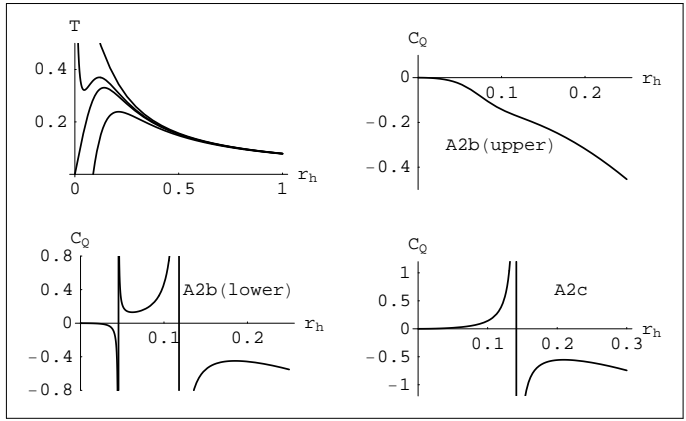}
\caption{Behaviour of $T(r_h,Q)$ and $C_{Q}(r_h,Q)$ as functions of $r_h$ for several values of $Q$ corresponding to the different \textbf{A2} cases. This plot is obtained from the BI model, as representative of the small-$r_{h}$ behaviour of the \textbf{A2} models with $\sigma > 2$. For the temperature (upper left panel) there are two \textbf{A2b} cases (two upper curves, with $Q < Q_{c}$), one of them exhibiting two local extrema ($Q\lesssim Q_C$) and the other, obtained for lower values of $Q$, behaving nearly as the Schwarzschild BH temperature. The limit $r_h = 0$ corresponds to the divergent-temperature non-extreme BPs. The curve starting at the origin corresponds to the critical value of the charge and exhibits a EBP there (case \textbf{A2c}). The lower curve, corresponding to $Q > Q_{c}$ (\textbf{A2a} cases) exhibits a behaviour similar to that of the \textbf{UVD} and \textbf{A1} cases. The associated specific heats are plotted in the other panels: upper right and lower left panels correspond to the two \textbf{A2b} temperature curves. The lower right panel corresponds to the cases \textbf{A2c}, with the curve starting at $r_{h} = 0$, but it is also representative of the cases \textbf{A2a} if the curve starts on a finite value of $r_{h}$, corresponding to the EBH configurations. The vertical asymptote of $C_{Q}$ correspond to local extrema of the temperature.}
\label{figure14}
 \end{center}
\end{figure}

Figure \ref{figure14} shows the behaviour of the temperature, for these BI black holes, as a function of $r_{h}$ for several values of the charge, corresponding to the cases \textbf{A2b} (two upper curves), \textbf{A2a} (lower curve) and \textbf{A2c}. We have also plotted the behaviour of the specific heats for the two \textbf{A2b} cases and the \textbf{A2c} case.

The \textbf{A2} models with $\sigma = 1$ lead to new thermodynamic behaviours. In particular, the temperature of the extreme black points becomes now finite. Let us give two examples of these cases, which are constructed by convenience, since they are illustrative of the different statements on the state variables discussed in section \ref{sectionIV}. The first one corresponds to a family of NEDs defined through the expression of their ESS solutions, which take the form

\be
E(R,Q) = \frac{a - b_{0} R + c_{0} R^{\lambda}}{1 + R^{\mu}},
\label{eq:(4-47)}
\en
($R = r/\sqrt{Q}$) where the exponents $\lambda > 1$ and $\mu > \lambda + 1$, and the positive constants $a$, $b_{0}$ and $c_{0}$ are to be chosen in such a way that $E(R)$ be positive and monotonically decreasing everywhere, as required for admissibility. From Eq.(\ref{eq:(4-47)}) and the first integral (\ref{eq:(2-4)}) we can obtain, through a quadrature, the expression of the admissible Lagrangian densities supporting these soliton-like elementary solutions. We shall make the choice $a = 1$, $b_{0} = 0.2$, $c_{0} = 0.1$, $\mu = 9/2$ and two different choices for the exponent $\lambda$ ($\lambda = 3/2$, corresponding to a \textbf{B3} asymptotic behaviour or $\lambda = 5/2$, corresponding to a \textbf{B2} asymptotic behaviour). Both choices lead to admissible models, for which the coefficients of the expansion of $E(R)$ around $R = 0$ coincide with those of the numerator of (\ref{eq:(4-47)}). The two choices for the exponent $\lambda$ lead to the same value of the extreme BP temperature but to different behaviours for the temperature of small-radius BHs (see table \ref{table:II}). For $\lambda = 3/2 < 2$ the temperature decreases from its BP value ($= 2 b_{0}/\sqrt{\pi a}$) with infinite negative slope whereas for $\lambda = 5/2 > 2$ the temperature increases from the same value at $r_{h} = 0$ with finite slope ($= 2\Delta)$. In both cases the specific heats of the extreme BP configurations vanish, as

\be
C_{Q}(r_{h} \rightarrow 0,Q_{c}) \sim -\frac{\pi^{3/4} b_{0}}{c_{0} a^{1/4}} r_{h}^{3/2} = -2\pi^{3/4} r_{h}^{3/2},
\label{eq:(4-48)}
\en
for the model with $\lambda = 3/2$, and as

\be
C_{Q}(r_{h} \rightarrow 0,Q_{c}) \sim \sqrt{\frac{\pi}{a}} \frac{2b_{0}}{\Delta}r_{h} = \frac{2\sqrt{\pi}}{5\Delta} r_{h},
\label{eq:(4-49)}
\en
for the model with $\lambda = 5/2$.

\begin{figure}[h]
 \begin{center}
\includegraphics[width=0.75\textwidth]{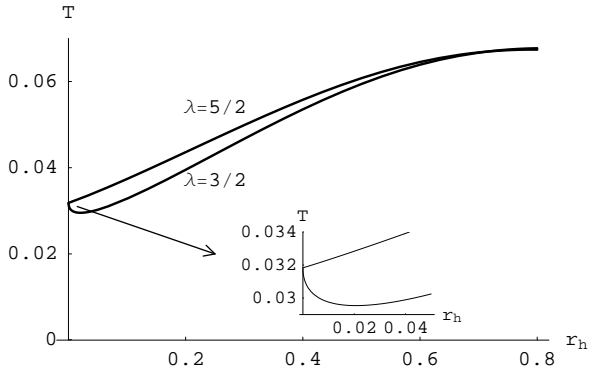}
\caption{Behaviour of $T(r_h,Q)$ at constant $Q = Q_{c}$ and for small radius BHs associated to the fields (\ref{eq:(4-47)}) for two choices of the parameter $\lambda=5/2$ (upper curve) and $\lambda=3/2$ (lower curve) with fixed values for the other parameters (see the text). Note the qualitative difference in the behaviour of the EBPs, which exhibit both the same finite temperature but with a finite positive slope for $\lambda=5/2$ and a negative divergent slope for $\lambda=3/2$. The small frame expands the region of configurations close to the EBPs.}
\label{figure15}
 \end{center}
\end{figure}

Figure \ref{figure15} shows the behaviour of the temperature as a function of $r_{h}$ at fixed $Q = Q_{c}$ for both models. Even though this behaviour is similar for finite horizon radius, the qualitative difference around $r_{h} = 0$ (extreme BP configurations) is apparent.

Finally, let us consider a family of models representative of the subclass of the \textbf{A2} NEDs satisfying the condition $\sigma = 1$, $\lambda = 3$ and $\Delta = 0$. The assumed form of the Lagrangian for $X \geq 0$ is

\be
\varphi(X,Y=0) = \frac{2}{\alpha - \sqrt{X}} + \frac{2}{\alpha^{2}}\left(\frac{2}{\alpha^{2}} X^{3/2} - X^{1/2} - \alpha \right),
\label{eq:(4-50)}
\en
where $\alpha$ is a positive constant parameterizing the family. It is easy to verify that, as a function of $X$, at $Y=0$, this expression increases monotonically from $\varphi(0,0) = 0$ and exhibits a vertical asymptote at $X = \alpha$. Its slope at the origin takes the finite value

\be
\frac{\partial \varphi}{\partial X}\Big \vert_{X \rightarrow 0} \sim \frac{2}{\alpha^{3}},
\label{eq:(4-51)}
\en
and, consequently, this is a one-parameter family of admissible models belonging to the class \textbf{A2-B2} (obviously, for $X < 0$ these Lagrangians should be properly extended taking into account the admissibility conditions). The general ESS solution of these models can be obtained explicitly in terms of the parameter $\alpha$ and the charge $Q$ by using the first integral (\ref{eq:(2-4)}) and reduces to solve a quartic algebraic equation. Nevertheless, we are interested here in the behaviour of the field and the thermodynamic functions near the extreme BP configurations. By replacing the expression (\ref{eq:(4-17)}) and the derivative of (\ref{eq:(4-50)}) in the first integral we determine straightforwardly the parameters of the first terms of the polynomial expansion of the field around the center as

\be
a = \alpha \hspace{0.1cm}; \hspace{0.1cm} b_{0} = 1\hspace{0.1cm}; \hspace{0.1cm} c_{0} = -\frac{5}{2\alpha^{2}}\hspace{0.1cm}; \hspace{0.1cm} \sigma = 1\hspace{0.1cm}; \hspace{0.1cm}\lambda = 3,
\label{eq:(4-52)}
\en
or, equivalently,

\be
E(r_{h} \rightarrow 0,Q) \sim \alpha - R_{h} - \frac{5}{2\alpha^{2}} R^{3}_{h},
\label{eq:(4-53)}
\en
($R_{h} = r_{h}/\sqrt{Q}$) and we see that $\alpha$ must be interpreted as the maximum field strength. Moreover, using Eq.(\ref{eq:(4-18)bis}) we are led to the value $\Delta = 0$ for the integration constant of the expansion of the Lagrangian around $r_{h} = 0$. Consequently, a glance to table \ref{table:II} shows that the temperature of the extreme BP configuration in these cases is $T_{EBP} = 2/\sqrt{\pi \alpha}$, the slope of the temperature at $r_{h} = 0$ vanishes and becomes negative as $r_{h}$ increases, and the specific heats attain, for the extreme BP configurations, the negative value $C_{Q;EBP} = -\alpha^{2}/40$.

\begin{figure}[h]
 \begin{center}
\includegraphics[width=0.75\textwidth]{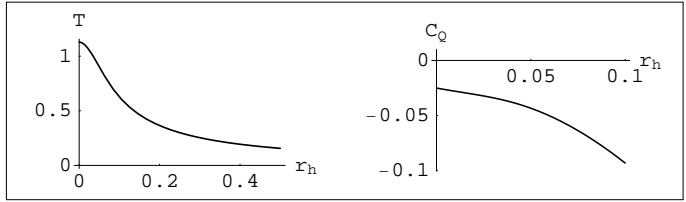}
\caption{Behaviour of $T(r_h,Q_C)$ and $C_Q(r_h,Q_C)$ for the model (\ref{eq:(4-50)}). As in the case of figure \ref{figure15} the temperature of the EBP configuration is finite, but now its slope vanishes there.}
\label{figure16}
 \end{center}
\end{figure}

Figure \ref{figure16} displays these temperature and specific heat behaviours for the critical ($Q=Q_{c}$) configurations, which become atypical in the limit of extreme BP configurations.

\section{Finite thermodynamic relations, scaling group and generating equations} \label{sectionV}

Let us consider now some finite relations arising between the state functions introduced in the precedent section.

\subsection{Generalized Smarr law}

In recalculating the derivative of $\varepsilon_{ex}$ with respect to the charge at constant $r_{h}$ in equation (\ref{eq:(4-4)}) we can use the scaling law of the external energy function (\ref{eq:(2-16)bis}) and we are led to the relation

\be
Q\Phi + TS = \frac{3M - r_{h}}{2},
\label{eq:(5-1)}
\en
or, alternatively,

\be
M = \sqrt{\frac{S}{9\pi}} + \frac{2}{3} (TS + Q\Phi),
\label{eq:(5-2)}
\en
where the dependence in $r_{h}$ has been replaced by dependence in the entropy $S$. In the particular case of the RN black hole this expression boils down to

\be
M = 2TS + Q\Phi,
\label{eq:(5-3)}
\en
which is the well known Smarr formula \cite{smarr73}. As another example, for the NED family defined in Ref.\cite{hassaine09-NED} by the Lagrangian density $\varphi(X,Y=0)=X^{p}$ (particularized here to three space dimensions), $p$ being an integer number, Eq.(\ref{eq:(5-2)}) provides the result

\be
M = 2TS + \frac{Q\Phi}{p},
\label{eq:(5-3)b)}
\en
which is precisely the Smarr-like law obtained in that reference.

Moreover, it is straightforward to show that the equivalence of (\ref{eq:(5-2)}) and (\ref{eq:(5-3)}), which requires

\be
V(r_{h},Q) = Q\Phi(r_{h},Q) - 2\varepsilon_{ex}(r_{h},Q) = 0,
\label{eq:(5-4)}
\en
arises only for the ESS black hole solutions of the Einstein-Maxwell model. This condition is related to the correspondence between the energy content of the field created by a system of charges at rest and their electrostatic potential energy, which holds in flat-space Maxwell electrodynamics \cite{landau}, but not for general NEDs. Thus, for any admissible G-NED the function $V(r_{h},Q)$ gives the deviations from the exact Smarr law, induced by the effect of the nonlinear self-couplings of the electric field. Owing to the scaling properties (\ref{eq:(4-10)}) and (\ref{eq:(2-16)bis}) this function scales as the external energy

\be
V(r_{h},Q) = Q^{3/2} V(R_{h},Q=1),
\label{eq:(5-5)}
\en
($R_{h} = r_{h}/\sqrt{Q}$) and is fully determined by its characteristic expression for $Q=1$.

We conclude that Eq.(\ref{eq:(5-2)}) is the natural generalization for admissible G-NEDs of the Smarr formula (\ref{eq:(5-3)}) of the Einstein-Maxwell theory.

\subsection{Other finite relations}

As already mentioned, the thermodynamical approach regards the ESS black hole solutions obtained from a given gravitating NED as different states of a unique system characterized by two state variables, the most immediate ones being the integration constants $M$ and $Q$, which arise in the resolution of the Einstein equations (\ref{eq:(3-3)}). On the other hand, in solving the thermodynamic problem for a given model, we start with the first integral of the electromagnetic field equations (\ref{eq:(2-4)}) and we can obtain the state variables as functions of $r_{h}$ and $Q$. New state variables can be defined as $\alpha = \alpha(r_{h},Q)$ and $\beta = \beta(r_{h},Q)$, where these functions are assumed to be invertible in some pertinent ranges, leading to

\be
r_{h} = f(\alpha,\beta) \hspace{.1cm}; \hspace{.1cm} Q = g(\alpha,\beta),
\label{eq:(5-6)}
\en
where the functions $f$ and $g$ can be multi-valuated.

The universal scaling laws obtained in section \ref{sectionIV} allow to build new finite relations between different thermodynamic variables, whose form is now dependent on the particular model. Let us write these scaling laws in a new (but equivalent) form for the main state variables introduced so far. First of all, the scaling law (\ref{eq:(2-4)bis}) for the electrostatic field leads to

\be
E(\theta r_{h},\theta^{2} Q) = E(r_{h},Q),
\label{eq:(5-6)bis}
\en
where $\theta$ is any positive parameter (which can be a state variable). From Eq.(\ref{eq:(4-16)bis}) we obtain for the temperature

\be
T(\theta r_{h},\theta^{2} Q) = \frac{1-\theta^{2}}{4\pi \theta r_{h}} + \theta T(r_{h},Q).
\label{eq:(5-7)}
\en
From Eq.(\ref{eq:(4-11)bis}) we obtain for the mass

\be
M(\theta r_{h},\theta^{2} Q) = \frac{\theta(1-\theta^{2})}{2} r_{h} + \theta^{3} M(r_{h},Q),
\label{eq:(5-8)}
\en
whereas for the electrostatic potential on the horizon we obtain, from Eq.(\ref{eq:(4-10)}),

\be
\Phi(\theta r_{h},\theta^{2} Q) = \theta \Phi(r_{h},Q).
\label{eq:(5-9)}
\en
Let us denote as $\Gamma(\theta)$ ($\theta>0$) these scaling transformations. From Eqs.(\ref{eq:(5-6)bis})-(\ref{eq:(5-9)}) it is easy to show that the product of two transformations satisfies

\be
\Gamma(\theta_{1}) \ast \Gamma(\theta_{2}) = \Gamma(\theta_{1} \cdot \theta_{2}).
\label{eq:(5-9)bis}
\en
Consequently the set of scaling transformations for the thermodynamical state functions associated to the BH solutions of any admissible G-NED has the structure of a multiplicative one-parameter group.

From these laws and the identification of $\theta$ with different state variables we can obtain several finite relations whose generic forms are the same for all models, but involve particular expressions associated to each NED. Let us give two simple examples. Consider first the function $\Phi(r_{h},Q)$ obtained from the first integral (\ref{eq:(2-4)}) (after a quadrature) for a particular model. Using the scaling law (\ref{eq:(5-9)}) with $\theta = T$ we are led to

\be
T\Phi = \Phi\left(T \sqrt{\frac{S}{\pi}},QT^{2}\right),
\label{eq:(5-10)}
\en
where $r_{h}$ has been written in terms of the entropy. In the same way, the explicit expression of the temperature (\ref{eq:(4-12)}) and the scaling law (\ref{eq:(5-7)}), with $\theta = \Phi$, lead to

\be
T\Phi = T\left(\Phi \sqrt{\frac{S}{\pi}},Q\Phi^{2}\right) + \frac{\Phi^{2}-1}{4 \Phi \sqrt{\pi S}}.
\label{eq:(5-11)}
\en
These equations, together with the generalized Smarr law (\ref{eq:(5-2)}), allow for the elimination of two of the five state variables $\Phi, Q, M, T$ and $S$, leading to different expressions for the ``equation of state" of the set of BHs associated to a given model.

\subsection{Generating equations}

The universality of the scaling laws allows for a more direct and enlightening procedure in generating the equations of state of general BH configurations. These laws lead to finite relations involving the derivatives of the state functions and are also ``universal", in the sense that they are first-order differential equations whose form is independent of the particular gravitating NED. The characteristic curves of these equations generate, through properly chosen boundary conditions, surfaces in the space of state variables involved in the particular scaling law. The points of each one of these surfaces correspond to BH states associated to a given gravitating NED. As the structure of the beam of characteristic curves is also universal, this procedure allows for a separate analysis of some model-independent thermodynamic properties and those model-dependent behaviours induced by the particular boundary conditions corresponding to each NED. Let us present here a particular example illustrating the method and some of the expected results.

\begin{figure}[h]
 \begin{center}
\includegraphics[width=0.75\textwidth]{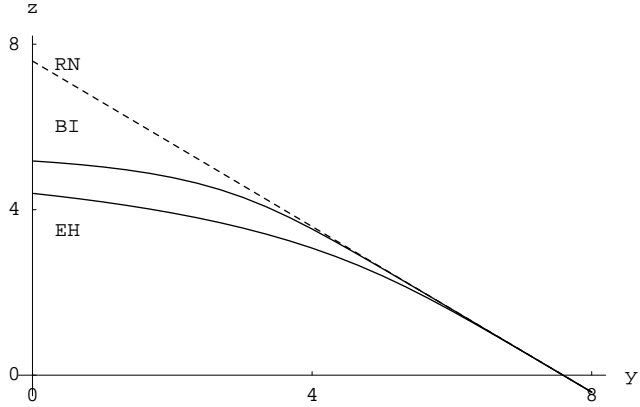}
\caption{$z=\ln(\Phi^2)$ versus $y=\ln(S)$ curves in Eq.(\ref{eq:(5-14)}) for the RN solution (dashed line) as compared to the BI and EH solutions, in solid style for some choices of the parameters and unit charge.}
\label{figure16b}
 \end{center}
\end{figure}

Using the expression (\ref{eq:(4-11)}) of the $Q$-derivative of the electrostatic potential at constant $r_{h}$ and replacing the electrostatic field on the horizon in terms of the gradient of $\Phi$ we are led to

\be
2Q \frac{\partial \Phi}{\partial Q}\Big\vert_{r_{h}} = \Phi - r_{h} \frac{\partial \Phi}{\partial r_{h}}\Big\vert_{Q}.
\label{eq:(5-12)}
\en
This equation (which can also be obtained directly by derivation of Eq.(\ref{eq:(5-9)}) with respect to the parameter $\theta$ at $\theta = 1$) involves the function $\Phi(r_{h},Q)$ and its first derivatives with respect to these arguments. It is remarkable because its ``universality", since it must be satisfied by the solutions of any G-NED, as a direct consequence of the scaling law (\ref{eq:(5-9)}), which is a universal property for field theories associated to the action (\ref{eq:(3-1)}). On the other hand, it is independent of the general laws of thermodynamics, as can be easily checked for other thermodynamical systems. Thus, the particular forms of the ``equations of state" of the systems of ESS black hole solutions associated to the different particular models, characterized by the two independent parameters $r_{h}$ and $Q$, are solutions of this equation, corresponding to surfaces in the $\Phi-r_{h}-Q$ space. In order to explore further this interpretation, let us write Eq.(\ref{eq:(5-12)}) under the form

\be
\frac{2Q}{\Phi} \frac{\partial \Phi}{\partial Q}\Big\vert_{S} + \frac{2S}{\Phi} \frac{\partial \Phi}{\partial S}\Big\vert_{Q} - 1 = 0,
\label{eq:(5-13)}
\en
where the horizon radius has been replaced by the entropy as independent variable. This expression suggests the introduction of the new variables

\be
x = \ln(Q), \hspace{.1cm} y = \ln(S), \hspace{.1cm} z = \ln(\Phi^{2})
\label{eq:(5-14)}
\en
in terms of which the equation takes the form

\be
\frac{\partial z}{\partial x} + \frac{\partial z}{\partial y} - 1 = 0.
\label{eq:(5-15)}
\en
This is a very simple first-order, linear, partial differential equation whose set of characteristics is the beam of straight lines parallel to the vector $(1,1,1)$ in the $(x,y,z)$ space. Consequently, the equations of state of the system of ESS black hole solutions associated to the G-NEDs are represented in this space by ruled surfaces generated by lines of this beam. In order to choose the particular surface associated to the system of BH solutions of a particular G-NED we can solve, for example, the first integral (\ref{eq:(2-4)}) for the characteristic ($Q = 1$) case and determine the form of $\Phi(r,Q=1)$ through a quadrature. This defines a curve $z = z(x=0,y)$ in the $y-z$ plane, which is a boundary condition for Eq.(\ref{eq:(5-15)}) (see figure \ref{figure16b}) determining a unique surface solution $z = z(x,y)$ in the $(x,y,z)$ space, generated by the beam of characteristics. In the $(\Phi,Q,S)$ space the parametric equations for the beam of characteristics of (\ref{eq:(5-13)}) read

\be
Q = Q_{0} \tau \hspace{.1cm}; \hspace{.1cm} S = S_{0} \tau \hspace{.1cm};\hspace{.1cm} \Phi = \Phi_{0} \sqrt{\tau},
\label{eq:(5-16)}
\en
where $\tau$ is the (positive) parameter and $Q_{0}, S_{0}$ and $\Phi_{0}$ are the coordinates of the points defining the different curves of the beam, corresponding to a particular BH configuration. These curves are parabolae having the origin as a common vertex, where the beam is singular.

A similar analysis can be carried out from the scaling laws of other state variables. For example, the scaling law for the temperature (\ref{eq:(5-7)}) leads to the generating equation

\be
\frac{\partial T}{\partial x} + \frac{\partial T}{\partial y} - \frac{T}{2} + \frac{e^{-y/2}}{\sqrt{\pi}} = 0,
\label{eq:(5-17)}
\en
where $x$ and $y$ are the same independent variables defined in Eq.(\ref{eq:(5-14)}). This is a linear, inhomogeneous, first-order, partial differential equation, generating the state surfaces $T = T(x,y)$ for the different G-NEDs, once the corresponding boundary conditions are specified. In the same way, the scaling law for the BH mass (\ref{eq:(5-8)}) leads to the generating equation

\be
\frac{\partial M}{\partial x} + \frac{\partial M}{\partial y} - \frac{3}{2}M + \frac{e^{y/2}}{\sqrt{\pi}} = 0,
\label{eq:(5-18)}
\en
whose set of solutions contains the state surfaces $M = M(x,y)$ associated with all the admissible G-NEDs.

These preliminary results require a more detailed investigation of the structures which lie at the ground
of these scaling symmetries and their consequences for the BH thermodynamics. A systematic exploration of this issue is in progress, but goes beyond the scope of this paper.

\section{Conclusions and perspectives} \label{sectionVI}

In this paper we have performed a generalization of the well known thermodynamic analysis of the RN black hole solutions of the Einstein Maxwell field equations to the BH solutions of any admissible gravitating NED. The exhaustive classification of the admissible NEDs in terms of the central and asymptotic behaviours of their elementary solutions in flat space, and the analysis of the geometric structure of their gravitating counterparts, have been carried out in references \cite{dr10a} and \cite{dr10b}. There, the asymptotic flatness of any gravitating ESS solution of these \textbf{admissible} NEDs has been established, regardless of the asymptotic behaviour of their integral of energy in flat space. When this integral diverges at large $r$ (always slowly than $r$, owing to the admissibility conditions) we are led to the families of asymptotically \textbf{IRD} models, whose gravitating BH solutions exhibit geometric structures similar to those of the other models at finite radius, but approach flatness at large $r$ slower than the Schwarzschild field. For such asymptotically anomalous models the standard thermodynamic analysis is meaningless, at least in the usual way (undefined ADM masses and asymptotic divergence of the electrostatic potentials).

For models whose ESS solutions have asymptotically convergent integrals of energy in flat space (\textbf{B} models) the gravitating elementary solutions behave as the Schwarzschild field at large $r$, no matter the asymptotic degree of convergence of these integrals. In such cases the ADM mass of the BH solutions is well defined, the electrostatic potential can be set to zero asymptotically and the thermodynamic laws can be consistently established. The thermodynamic properties of the BH solutions in these cases have been analyzed in terms of the behaviours of the integral of energy and the field around the center of the flat space ESS solutions. When this integral (and, hence, the field) diverge as $r \rightarrow 0$ (\textbf{UVD} models) the geometric structures (two-horizons BHs, single-horizon extreme BHs or naked singularities) and the associated thermodynamical behaviours are qualitatively similar to those of the Reissner Nordstr\"om BHs, although the temperature and the specific heats as functions of the horizon radius (or of the mass) can exhibit more involved behaviours, related to the more or less involved structure of $E(r)$ in each case. For models exhibiting finite-energy ESS solutions in flat space and whose electrostatic field diverges at the center (slower than $r^{-1}$, \textbf{A1} models) the associated BHs can exhibit new geometric features, as compared with the \textbf{UVD} cases (mainly the existence of non-extreme single-horizon BHs), but the form of the mass-radius relations and the behaviours outside the event horizon are qualitatively similar in both cases and hence we are led to similar thermodynamic properties.

For models whose ESS solutions in flat space are finite-energy with finite maximum field strength (\textbf{A2} family, to which the Born-Infeld model belongs) the set of associated gravitating ESS solutions exhibits a richer structure, including extreme and non-extreme black points as new possible configurations. These new configurations have thermodynamic properties which are related to the behaviour of the electrostatic field around the center. This relation has been extensively analyzed and, in particular, we have determined the conditions to be satisfied by these \textbf{A2} models in order to support extreme black point solutions with zero, finite or divergent temperature. Concerning the non-extreme black points, their temperature is always divergent.

We have performed a detailed analysis of the properties of the extremal BH solutions for the different families. The thermodynamic behaviour of these configurations differ strongly among the different families and suggest the possible existence of models exhibiting van der Waals-like first-order phase transitions in the $\Phi-Q$ plane.

Once the possible structures and thermodynamic properties of the BH solutions of the different families of admissible NEDs have been determined and classified, the results of the present analysis allow for the explicit construction of specific models whose solutions exhibit prescribed behaviours. In this way we have built some new models as examples illustrating different aspects of the study carried out here.

From the invariance of scale of the general NED field equations in flat space, we have established universal scaling laws, satisfied by the thermodynamic functions defining the BH states of any model. These scaling laws lead to universal first-order partial differential equations involving sets of three state variables, whose solutions are surfaces in the three-dimensional spaces defined by these variables. Each one of these surfaces is generated by the beam of characteristics associated to the equation. From appropriate boundary conditions we can select the surface solution associated to a particular G-NED. The points of this surface correspond to the BH states which are the solutions of this model, and its analytic equation, relating the involved variables, defines a expression of the ``equation of state" of this system of BHs.

To conclude, let us mention some perspectives and possible extensions of this study:

First, in Ref.\cite{dr09} we have established that the analysis of the ESS problem for NEDs in flat space can be extended to the same problem for generalized non-Abelian gauge field theories of compact semi-simple Lie groups. More explicitly, we have shown that the ESS solutions of a generalized gauge field theory, defined by a Lagrangian density of the form (\ref{eq:(2-1)}) (in terms of the usual gauge invariants with the ordinary trace definition in this non-Abelian case) have the same functional dependence in $r$ as those of the NED defined by a Lagrangian density of the same form (in terms of the Abelian gauge invariants). In this way we have recently communicated some preliminary results on the extension of the analysis performed here to systems of non-rotating BHs of non-Abelian origin \cite{dr11}. A more detailed report on this issue is in progress.

Second, the extensions of the present general analysis (geometric structures and thermodynamics) to the cases of NEDs coupled to more general actions for the gravitational field (e.g. with a cosmological constant term, or Gauss-Bonnet and Lovelock theories) is in progress, a context where some particular cases of NEDs have been already explored in the literature \cite{aiello}.

Finally, let us stress that a more complete analysis of the scaling symmetries and their consequences on the BH thermodynamics is necessary. In particular, the study of the structures of the beams of characteristics associated to different generating equations and of the boundary conditions selecting the equations of state of the BH systems associated with different NEDs, allows for the exploration of eventual phase transitions, an issue also under development.

\begin{acknowledgements}
We are indebted to Mr. Charles Horeau for useful discussions. This work has been
partially funded by the project MICINN-09-FPA2009-11061 and grants MEC-12-PR2011- 0334 and
AMCEI-UO-2012/2013 (Spain).

\end{acknowledgements}

\end{document}